\documentclass[
    reprint,
	amsmath,
    amssymb,
	aps,
	prx,
    floatfix,
    superscriptaddress
]{revtex4-2}

\usepackage[english]{babel}
\usepackage[utf8x]{inputenc}
\usepackage[T1]{fontenc}
\usepackage{braket}
\usepackage{mathtools}
\usepackage{flushend}

\usepackage{amsmath}
\usepackage{siunitx}
\usepackage{graphicx}
\usepackage{dcolumn}
\usepackage{bm}
\usepackage{array}
\usepackage[shortlabels]{enumitem}
\newcolumntype{?}{!{\vrule width 1pt}}
\usepackage[colorinlistoftodos]{todonotes}
\usepackage[colorlinks=true,linkcolor=blue,citecolor=blue,urlcolor=blue]{hyperref}
\usepackage{slashed}

\usepackage{bibunits}
\defaultbibliographystyle{apsrev4-1}
\usepackage{etoolbox}

\makeatletter
\newcommand*{\newbibstartnumber}[1]{%
  \apptocmd{\thebibliography}{%
    \global\c@NAT@ctr #1\relax
    \addtocounter{NAT@ctr}{-1}%
  }{}{}%
}
\makeatother

\newcommand\bb[1]{\mbox{\boldmath{$#1$}}}
\newcommand{\msb}[1]{\bb{\mathsf{#1}}}
\newcommand\grad{\bb{\nabla}}
\newcommand\bcdot{\,\bb{\cdot}\,}

\newcommand\rmd{{\rm d}}

\def\apj{Astrophys.~J.}
\def\apjl{Astrophys.~J.~Lett.}
\def\aap{Astron.~Astrophys.}
\def\araa{Ann.~Rev.~Astron.~Astrophys.}

\def\jopp{J.~Plasma Phys.}
\def\mnras{Mon.~Not.~R.~Astron.~Soc.}

\def\natp{Nature Phys.}
\def\njp{New J.~Phys.}
\def\pnas{Proc.~Nat.~Acad.~Sci.}

\def\pre{Phys.~Rev.~E}
\def\prl{Phys.~Rev.~Lett.}
\def\prr{Phys.~Rev.~Res.}

\def\physrep{Phys.~Rep.}
\def\pop{Phys.~Plasmas}
\def\pof{Phys.~Fluids}

\def\rpp{Rep.~Progress Phys.}
\def\ssrv{Space Sci.~Rev.}

\makeatletter
\let\cat@comma@active\@empty
\makeatother
\begin{document}
\title{Spontaneous magnetization of collisionless plasma through the action of a shear flow}
\author{Muni Zhou}
\thanks{\href{mailto:munizhou@mit.edu}{munizhou@mit.edu} \vspace{0.5cm}}
\affiliation{Plasma Science and Fusion Center$,$ Massachusetts Institute of Technology$,$ Cambridge$,$ MA 02139$,$ USA}
\author{Vladimir Zhdankin}
\affiliation{Center for Computational Astrophysics$,$ Flatiron Institute$, $ 162 Fifth Avenue$,$ NY 10010$,$ USA}
\affiliation{Department of Astrophysical Sciences$,$ Princeton University$, $ Peyton Hall$, $ Princeton$,$ NJ 08544$,$ USA}
\author{Matthew W.~Kunz}
\affiliation{Department of Astrophysical Sciences$,$ Princeton University$, $ Peyton Hall$, $ Princeton$,$ NJ 08544$,$ USA}
\affiliation{Princeton Plasma Physics Laboratory$,$ PO Box 451$,$ NJ 08544$,$ USA}
\author{Nuno F.~Loureiro}
\affiliation{Plasma Science and Fusion Center$,$ Massachusetts Institute of Technology$,$ Cambridge$,$ MA 02139$,$ USA}
\author{Dmitri A.~Uzdensky}
\affiliation{Center for Integrated Plasma Studies$,$ Physics Department$,$ UCB-390$,$ University of Colorado$,$ Boulder$,$ CO 80309$,$ USA}

\date{\today}

\begin{abstract}
We study in a fully kinetic framework the generation of seed magnetic fields through the Weibel instability driven in an initially unmagnetized plasma by a large-scale shear force. We develop an analytical model that describes the development of thermal pressure anisotropy via phase mixing, the ensuing exponential growth of magnetic fields in the linear Weibel stage, and its saturation when the seed magnetic fields become strong enough to instigate gyromotion of particles and thereby inhibit their free-streaming. The predicted scaling dependencies of the saturated seed fields on key parameters (e.g., ratio of system scale to electron skin depth, the forcing amplitude) are confirmed by 3D and 2D particle-in-cell simulations using an electron-positron plasma. This work demonstrates the spontaneous magnetization of a collisionless plasma through large-scale motions as simple as a shear flow, and therefore has important implications for magnetogenesis in dilute astrophysical systems.
\end{abstract}

\maketitle
\section{Introduction}

The origin and evolution of cosmic magnetism remains one of the most profound mysteries  in astrophysics and cosmology~\cite{widrow2002,kulsrud2008origin}.
Observations of Faraday rotation, Zeeman splitting, and synchrotron emission suggest pervasive ${\sim}\mu$G magnetic fields in our Galaxy and in the intracluster medium (ICM) of galaxy clusters~\cite{beck1996galactic,carilli2002cluster,beck2016magnetic}.  
It is widely believed~\cite{Arshakian2009,Ryu2012,donnert2018magnetic} that such dynamically important magnetic fields first arose as weak ``seed'' fields generated by cosmic batteries, subsequently amplified to currently observed levels by the turbulent dynamo---a fundamental plasma process that converts the mechanical energy of plasma motions into magnetic energy through electromagnetic induction. However, neither the origin problem---
what are the physical mechanisms underpinning these batteries---nor the dynamo problem---how magnetic fields are amplified and sustained by turbulent plasma motions---are well understood.

There are two broad perspectives on the origin of cosmic seed magnetic fields.
One suggests a primordial origin, whereby seed fields are generated by exotic early-Universe mechanisms during inflation or during cosmological phase transitions (e.g.,~\cite{grasso2001magnetic,widrow2012first,subramanian2016origin}). 
The other postulates an astrophysical origin, in which seed fields are generated by plasma processes occurring during structure formation and stellar evolution in the early Universe (e.g.,~\cite{kulsrud1997protogalactic,gruzinov2001gamma}). 
Famous examples of such plasma processes include the Biermann battery~\cite{biermann1950biermann}, which is thought to produce extremely weak (${\sim}10^{-20}~{\rm G}$) seed magnetic fields on macroscopic system scales~\cite{Pudritz1989,Subramanian1994,ryu1998,Gnedin2000}, and the Weibel instability~\cite{weibel1959,burton1959}, which can produce seed fields with near-equipartition strength but at microscopic plasma-inertial length scales.
As a plausible key ingredient of magnetogenesis~\cite{schlicheiser2003,Lazar2009}, the Weibel instability has been studied extensively in a variety of contexts such as collisionless shocks in both the relativistic~\cite{medvedev1999generation,silva2003interpenetrating,spitkovsky2008} and sub-relativistic~\cite{Kato2008,medvedev2006cluster} regimes, and in laser experiments~\cite{fox2013filamentation,huntington2015observation}.

Once formed, seed magnetic fields are thought to be amplified and sustained by the turbulent plasma dynamo. Previous dynamo studies---whether conducted within the framework of magnetohydrodynamics (MHD)~\cite{brandenburg2005astrophysical,schekochihin2004simulations,rincon2019dynamo} or, more recently, using a kinetic description~\cite{rincon2016turbulent,st2018fluctuation,pusztai2020dynamo}---assumed the existence of a seed field as an initial condition, and thus did not address its origin.
The possibility that, in a collisionless plasma (e.g., the intergalactic/intracluster medium), the turbulent motions of dynamo may themselves give rise to seed fields and thus magnetize the plasma non-inductively has not been adequately addressed. This idea presents intriguing questions that have not been addressed before; namely, how, exactly, are seed fields generated by generic large-scale motions? what are the strength and morphology of these self-consistently produced seed fields? can they seed the plasma dynamo, thereby yielding a fully self-consistent solution to the problem of magnetogenesis?

In this work, we aim to understand how an initially unmagnetized plasma may magnetize itself through kinetic instabilities arising self-consistently under the action of large-scale flows, which are ubiquitous and driven by a variety of large-scale processes in astrophysical environments. In unmagnetized, collisionless environments, the plasma flows are not of a pure fluid nature; instead, they are subject to phase mixing and Landau damping. 
As we will show, these cause the plasma distribution function to become anisotropic in velocity space, thereby providing free energy for microscopic instabilities such as Weibel to grow rapidly on top of the slowly varying macroscopic flows. The Weibel instability produces fluctuations that extract free energy from the thermal anisotropy and generate kinetic-scale ``seed'' magnetic fields. 
As the Weibel magnetic field grows, the plasma becomes magnetized, leading to the saturation of the instability and regulation of the macroscopic flows.

At plasma-kinetic scales, any macroscopic flow may be viewed locally as a shear flow and/or a compressional flow.
In this paper we focus on a shear flow and demonstrate its ability to spontaneously magnetize the plasma.
We adopt a fully kinetic framework in which the kinetic physics of both particle species is treated self-consistently.
The sequence of events through which the plasma becomes magnetized involves multiple stages, each of which we consider in detail. 
In Sec.~\ref{sec:theory}, we present our analytical model for each stage.
We then test this model using kinetic particle-in-cell (PIC) simulations, whose details are provided in Sec.~\ref{sec:setup} and from which the numerical results presented in Sec.~\ref{sec:numerical_results} are obtained.
We conclude in Sec.~\ref{sec:discussion} with a brief discussion of astrophysical implications and some thoughts on how our results fit into the broader narrative of cosmic magnetogenesis.

\section{Theory}
\label{sec:theory}
\subsection{Formulation of the problem and dimensionless parameters.}

Consider a three-dimensional (3D) system initialized with a uniform static Maxwellian plasma and negligible electromagnetic fields. 
The plasma has both negative and positive charges; a subscript $s$ is added to quantities to represent these two species ($s\in\{e,i\}$ for an electron-ion plasma and $s\in \{e,p\}$ for an electron-positron plasma).  
Each species is represented by its distribution function in phase space $f_s(t, \bb{x},\bb{v})$, 
mass $m_s$, and temperature $T_s$.
We limit our discussion to the sub-relativistic regime, in which the thermal and flow velocities of both species are much smaller than the speed of light $c$.
In this limit, the bulk flow velocity 
\begin{align}
    \bb{U}_s(t,\bb{x}) \equiv \biggl( \int\rmd^3\bb{v} \, \bb{v} f_s \biggr) \Big/ n_s(t,\bb{x}),
        \label{eq:define_Us}
\end{align}
where $n_s(t,\bb{x}) \equiv \int\rmd^3\bb{v}\ f_s$ is the density, and the thermal pressure tensor,
\begin{align}
    \msb{P}_s(t,\bb{x}) \equiv \int\rmd^3 \bb{v} \, m_s (\bb{v}-\bb{U}_s)(\bb{v}-\bb{U}_s) f_s,
    \label{eq:define_Ps}
\end{align}
are two basic quantities characterizing the bulk and thermal motions of the plasma, respectively.

In this initially unmagnetized static Maxwellian system, we consider a shear flow driven continuously by a time-independent external body force $\bb{F}_{\rm ext}(x)=m_s \bb{a}(\bb{x})$~\footnote{This study investigates the kinetic effects that spontaneously emerge on top of a large-scale shear flow. The only purpose of the external force is to provide such a macroscopic flow. To achieve this, we consider a gravity-type body force that leads to the same body acceleration $a_0$ for both species and drives a hydrodynamic flow. 
One can alternatively consider the body force applied with equal magnitude to both species. In this case, electrons will more readily respond to the force because of their smaller inertia, resulting in an electric current and electromagnetic fields. These detailed dynamics occur on the electron plasma-oscillation time scale and are not considered in this paper.
The choice between the same body acceleration or the same body force for the two species does not affect the comparison of our theory to the numerical simulations we performed, as the latter consider a pair plasma (in which case both approaches are equivalent).}.
The force is in the $\hat{\bb{y}}$ direction with a sinusoidal spatial variation in the $\hat{\bb{x}}$ direction, giving rise to a species-independent acceleration $\bb{a}(x) = a_0 \sin{(2\pi x/L)}\hat{\bb{y}}$, where $a_0$ is the constant amplitude of the acceleration and $L$ is the system scale. 

We define three time-dependent dimensionless parameters to represent the evolution of the system's energetics.
The first is the Mach number $M_s \equiv \sqrt{\braket{U_s^2}}/v_{{\rm th}s}$, where $U_s=|\bb{U}_s|$, $v_{{\rm th}s} \equiv \sqrt{T_s(t=0)/m_s}$ is the initial thermal speed, and $\braket{...}$ denotes a volume average. 
The Mach number squared $M_s^2 \approx \langle P_{{\rm bulk},s} \rangle/\braket{P_s}$, where $P_{{\rm bulk},s} \equiv m_s  n_s U_s^2$ is twice the bulk kinetic energy density (ram pressure) and $P_s \equiv n_s T_s \approx m_s n_s v^2_{{\rm th}s}$ is the thermal pressure of plasma, the latter approximation being accurate if the temperature $T_s$ does not change significantly over time.

The second dimensionless quantity is the thermal pressure anisotropy, $\Delta_s \equiv \sqrt{ \langle (P_{{\rm max},s}/P_{\perp,s})^2 \rangle}-1$, where $P_{{\rm max},s}$ is the maximum eigenvalue of the local thermal pressure tensor $\msb{P}_s$, and $P_{\perp,s}$ is the average of the other two eigenvalues associated with the two directions perpendicular to that of $P_{{\rm max},s}$.
Under the assumption of small pressure anisotropy ($P_s \approx P_{\perp,s}$), we have $\Delta_s \approx \braket{\Delta P_s}/\braket{P_s}$, where $\Delta P_s \equiv P_{{\rm max},s}-P_{\perp,s}$ represents the free energy density stored in pressure anisotropy.
Our definition of pressure anisotropy is different from the commonly used definition in terms of $P_\perp$ and $P_\parallel$ based on a preferred magnetic-field direction. 
In the absence of magnetic fields, we identify the local maximum thermal-pressure component and use it as a preferred direction.

Finally, the third dimensionless quantity is the inverse plasma beta, $\beta_s^{-1}$, where $\beta_s \equiv \braket{P_s}/\braket{B^2/8\pi}$ and $B(t,\bb{x})$ is the magnetic-field strength.
It represents the magnetic energy density normalized to the thermal pressure, and is thus the main quantitative characteristic we use to diagnose the growth of magnetic fields.
It is effectively zero when the magnetic field is initially negligible.
By analyzing the evolution of $M_s^2$, $\Delta_s$, and $\beta_s^{-1}$, we learn the energy partition amongst different energy reservoirs. 
In the following subsections, we describe distinct stages of the evolution as the system is continuously driven by the external shear force.

\subsection{Unmagnetized stage.} 
\label{sec:theory_unmagnetized}
In the initial, unmagnetized, stage, the electromagnetic fields are negligible. The system can thus be described by the following nonrelativistic Vlasov equation for each species, where the only acceleration is supplied by the external force~\footnote{No charge separation, and thus no electrostatic field, is expected if both species have the same body acceleration. Indeed, the solution for $f_s$ [Eq.~\eqref{eq:f_exact_uncharged_theory}] does not give rise to any charge separation, consistent with the assumption.}:
\begin{equation}
        \frac{\partial f_s}{\partial t} + v_x \frac{\partial f_s}{\partial x} + a_0 \sin\left(\frac{2\pi}{L} x\right) \frac{\partial f_s}{\partial v_y} = 0.
\label{eq:main_vlasov_theory}
\end{equation}
This unmagnetized system is 1D in position space and so the convective term, $\bb{v}\bcdot\grad f_s$, reduces to $v_x \partial_x f_s$. 
The exact solution of Eq.~\eqref{eq:main_vlasov_theory} can be obtained by the method of characteristics:
\begin{equation}
\begin{aligned}
    \label{eq:f_exact_uncharged_theory}
    f_s(t,x,&\bb{v})=f_{{\rm M},s}\left(\sqrt{v_x^2+v_z^2+ \widetilde{v_y}^2}  \right), \text{where}\\ 
    \widetilde{v_y} &\equiv v_y+
    \frac{L a_0}{2 \pi v_x}\left[\cos \left(\frac{2 \pi}{L} x \right)-\cos\left(\frac{2 \pi}{L} (x- v_x t) \right)\right].
\end{aligned}
\end{equation}
Here $f_{{\rm M},s}$ is the initial Maxwellian distribution for each species $f_{s}(0,x,\bb{v})=f_{{\rm M},s}(|\bb{v}|)\equiv n_{0s}/(\sqrt{2 \pi} v_{{\rm th}s})^3 \exp(-|\bb{v}|^2/2v_{{\rm th}s}^2)$, where $n_{0s}$ is the initial density. 
Under the normalization $\hat{t}=t v_{{\rm th}s}/L$, $\hat{\bb{v}}=\bb{v}/v_{{\rm th}s}$, $\hat{x}=x/L$, and $\hat{a}_0=a_0L/v_{{\rm th}s}^2$,  Eq.~\eqref{eq:main_vlasov_theory} can be reduced to the dimensionless form  $\partial_{\hat{t}}f_s+\hat{v_x}\partial_{\hat{x}}f_s+\hat{a}_0\sin(2\pi \hat{x})\partial_{\hat{v_y}}f_s=0$; this form shows that $\hat{a}_0$ is the only dimensionless free parameter controlling the overall dynamics.
In this solution, $f_s$ remains Maxwellian in $v_z$, and therefore one of the eigenvectors of the local pressure tensor $\msb{P}_s$ is fixed in the $z$-direction with its corresponding eigenvalue $P_{zz,s}$. The other two eigenvectors corresponding to the largest and smallest eigenvalues of $\msb{P}_s$, denoted as $P_{{\rm max},s}$ and $P_{{\rm min},s}$, are thus in the $x$-$y$ plane.

\begin{figure}
    \centering
    \includegraphics[width=0.45\textwidth]{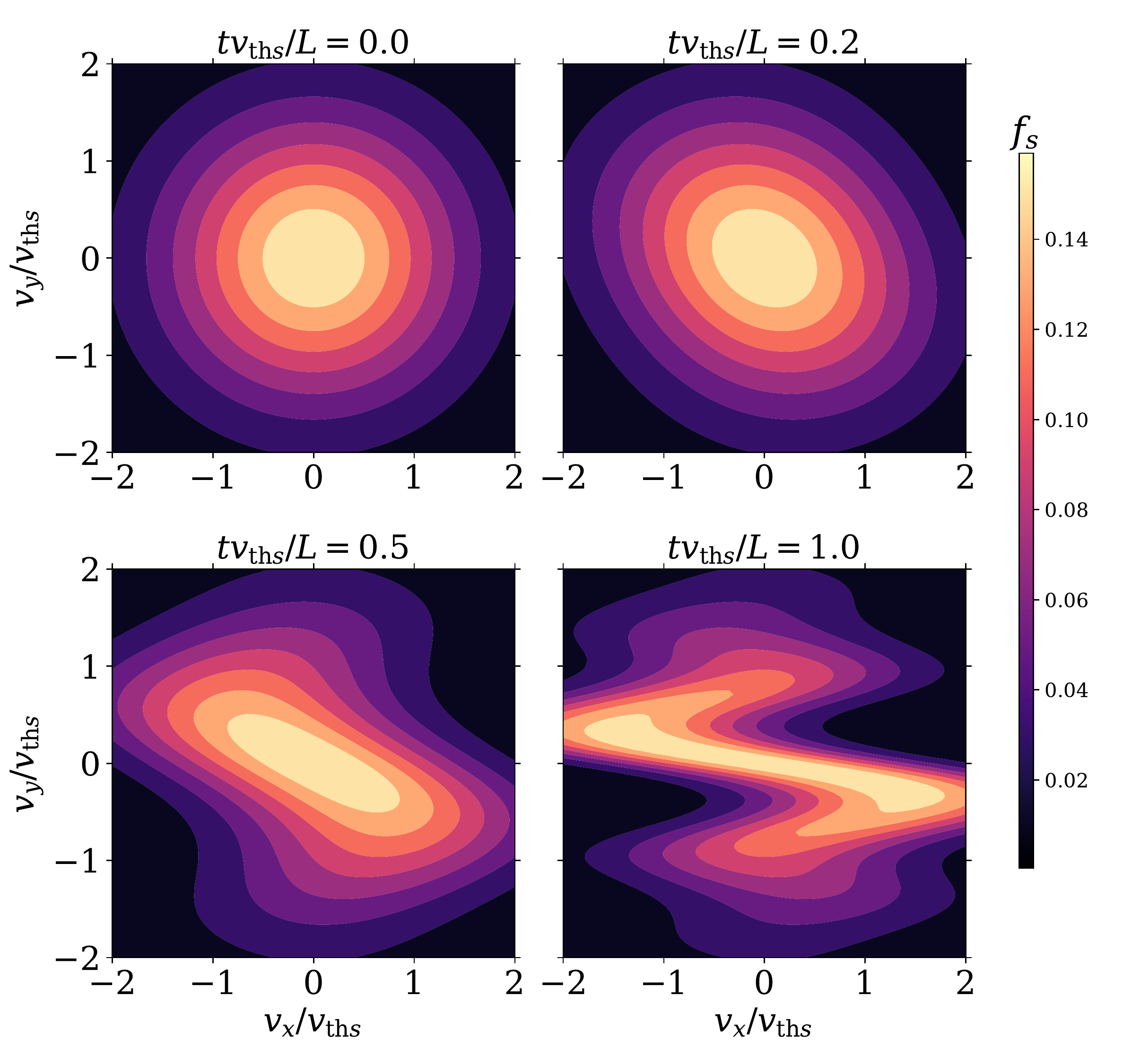}
    \caption{Contours of $f_s$ [Eq.~\eqref{eq:f_exact_uncharged_theory}] integrated over $v_z$ at different moments of time. The location $x=0$ with maximum shear is chosen and $\hat{a}_0=0.2 \pi^2$. The distribution is distorted by the phase mixing of momentum.}
    \label{fig:f0contour}
\end{figure}

In Fig.~\ref{fig:f0contour}, we show a visualization of the evolution of $f_s$ [Eq.~\eqref{eq:f_exact_uncharged_theory}] integrated over $v_z$ in the $v_x$-$v_y$ phase space for the choice $\hat{a}_0=0.2 \pi^2 \approx 2$. 
The solution is plotted at $x=0$, where the maximum shear occurs. 
The sinusoidal acceleration $\bb{a}$ gives rise to an $x$-dependent bulk flow $\bb{U}_s=U_s(t,x)\hat{\bb{y}}$. 
The transport of this non-uniform $y$-momentum is carried by particles streaming in the $x$-direction with their thermal speeds. 
This gives rise to the phase-mixing feature indicated in Fig.~\ref{fig:f0contour} by the distortion of $f_s$ in velocity space.  
The anisotropy developed in $f_s$ leads to the generation of thermal pressure anisotropy, $\Delta_s(t)$---a purely kinetic phenomenon which would be absent if the plasma were a collisional fluid.

We now proceed to calculate the evolution of $M_s(t)$ and~$\Delta_s(t)$. The time evolution of $\bb{U}_s(t,x)$ and $\msb{P}_s(t,x)$ can be calculated by taking moments of $f_s(t,x,\bb{v})$ [Eq.~\eqref{eq:f_exact_uncharged_theory}] following Eqs.~\eqref{eq:define_Us} and \eqref{eq:define_Ps}.
While their exact, finite-time expressions are not analytically integrable, we can take the second-order Taylor expansion of Eq.~\eqref{eq:f_exact_uncharged_theory} for $\epsilon \equiv tv_{{\rm th}s}/L \ll 1$ to obtain the early-time behavior:
\begin{equation}
\begin{aligned}
\label{eq:fs_2nd_order}
    f_s& (t,x,\bb{v})=f_{{\rm M},s}\left(|\bb{v}|\right) \Bigg\{ 1+ 
    \hat{a}_0 \frac{v_y}{v_{{\rm th}s}} \sin\Big(\frac{2 \pi}{L}x\Big) \frac{t v_{{\rm th}s}}{L} \\
    \mbox{} & -\frac{1}{2} \Bigg[2\pi \hat{a}_0 \frac{v_x v_y}{v_{{\rm th}s}^2} \cos\Big( \frac{2 \pi}{L}x \Big) \\ 
    \mbox{} &\quad +\hat{a}_0^2 \Big(1- \frac{v_y^2}{v_{{\rm th}s}^2} \Big) \sin^2 \Big( \frac{2 \pi}{L}x \Big) \Bigg] \left(\frac{t v_{{\rm th}s}}{L}\right)^2 \Bigg\} + \mathcal{O}(\epsilon^3).
\end{aligned}
\end{equation}
The first and second moments of Eq.~\eqref{eq:fs_2nd_order} provide the time evolution of the local bulk flow speed and local pressure anisotropy:
\begin{align}
    \label{eq:Us_x_2nd}
    \frac{U_s(t,x)}{v_{{\rm th}s}} &= \hat{a}_0 \sin \left(\frac{2\pi}{L}x \right) \frac{t v_{{\rm th}s}}{L}+ \mathcal{O}(\epsilon^3),\\
    \left( \frac{P_{{\rm max},s}}{P_{\perp,s}}-1 \right) (t,x)&= \frac{3\pi}{2}\hat{a}_0 \left|\cos \left(\frac{2\pi}{L}x \right)\right| \left( \frac{t v_{{\rm th}s}}{L} \right)^2 +\mathcal{O}(\epsilon^3),
    \label{eq:Delta_x_2nd}
\end{align}
respectively.
Up to second order in $\epsilon$, the bulk flow velocity is simply $U_s(x,t)\hat{\bb{y}} = \bb{a}(x) t$---identical to the fluid-level behavior for constant acceleration by a constant external force. The time evolution of $M_s(t)$ and $\Delta_s(t)$ can be obtained by calculating the root-mean-square values of Eqs.~\eqref{eq:Us_x_2nd} and \eqref{eq:Delta_x_2nd} over the domain:
\begin{align}
    \label{eq:Us_2nd}
    M_s(t) &=  \frac{1}{\sqrt{2}} \hat{a}_0 \frac{t v_{{\rm th}s}}{L} +\mathcal{O}(\epsilon^3),\\
    \Delta_s(t) &= \frac{3 \pi}{2\sqrt{2}} \hat{a}_0 \left( \frac{t v_{{\rm th}s}}{L} \right)^2 +\mathcal{O}(\epsilon^3).
    \label{eq:Delta_2nd}
\end{align}
Both $M_s$ and $\Delta_s$ increase on the thermal-crossing time scale, $L/v_{{\rm th}s}$, of their corresponding species.

Eqs.~\eqref{eq:Us_2nd} and \eqref{eq:Delta_2nd} are only valid on times short compared to the characteristic macroscopic time scale $L/v_{{\rm th}s}$. 
At later times, the Mach number $M_s$ asymptotes to a constant value; this occurs due to kinetic effects, namely, an effective kinetic viscosity, as we now describe. In the absence of the imposed shear flow, the particles in the collisionless, unmagnetized plasma that we consider would have an infinitely long mean free path.
However, in the presence of the shear flow, when particles travel a distance on the order of the characteristic length of the gradient of the shear flow ($L/2\pi$), the acceleration exerted on them changes sign and thus changes the direction of particle motion along the $y$-axis.
This is similar to a particle scattering process, setting an effective mean free path $\lambda_{\rm mfp} \simeq L/2\pi$, and giving rise to an effective viscosity $\nu_{\rm eff}\simeq v_{{\rm th}s}\lambda_{\rm mfp}$ for the fluid of both species.
The associated viscous force on the bulk flow, $\bb{F}_\nu(t,x) \simeq m_s \nu_{\rm eff} \nabla^2U_s(t,x) \hat{\bb{y}}$, is initially small but increases with $U_s$.
Eventually, it becomes comparable to the external force on the bulk fluid, $\bb{F}_\nu(t,x) \simeq \bb{F}_{\rm ext}(x)$, causing the bulk flow to stop accelerating.
This force-balance condition, combined with the estimation of $\lambda_{\rm mfp}$ and $\nu_{\rm eff}$, provides us with an estimate of the saturated characteristic bulk flow velocity $U_s^{\rm sat}$:
\begin{equation}
    m_s v_{{\rm th}s} \frac{L}{2\pi} \frac{U_s^{\rm sat}}{(L/2\pi)^2} \simeq m_s a_0.
    \label{eq:forcebalance}
\end{equation}
The saturated Mach number for each species can thus be written as
\begin{equation}
    M_s^{\rm sat} \equiv \frac{U_s^{\rm sat}}{v_{{\rm th}s}} \simeq \frac{a_0 L }{v_{{\rm th}s}^2 2\pi} = \frac{\hat{a}_0}{2\pi}.
    \label{eq:Msat}
\end{equation}
We denote by $\tau_0$ the moment of time that $M_s^{\rm sat}$ is reached, normalized to $L/v_{{\rm th}s}$; it is $\tau_0 \sim (U_s^{\rm sat}/a_0)/(L/v_{{\rm th}s})\sim (2\pi)^{-1}$. 
Eq.~\eqref{eq:Msat} suggests that the dimensionless parameter $\hat{a}_0$ represents a characteristic Mach number of the system during this stage of evolution.

Note that these estimates are predicated on the fact that the electromagnetic field remains negligible up until the saturation time. 
Realistically, however, the Weibel instability will be triggered by the developing pressure anisotropy and generate magnetic fields strong enough to magnetize the plasma on the kinetic time scale ${\sim} \omega^{-1}_{{\rm p}s} \ll L/v_{{\rm th}s}$, where $\omega_{{\rm p}s}$ is the plasma frequency for species $s$.
Therefore, the unmagnetized solution, Eq.~\eqref{eq:f_exact_uncharged_theory}, is only valid at very early times (on the fluid time scale) of the evolution $tv_{{\rm th}s}/L \ll 1$, during which the expressions for the time evolution of $M_s(t)$ and $\Delta_s(t)$ [Eqs.~\eqref{eq:Us_2nd} and \eqref{eq:Delta_2nd}] are good approximations.

\subsection{Linear Weibel stage.}
\label{sec:theory_linear}

As the pressure anisotropy increases [Eq.~\eqref{eq:Delta_2nd}], an electromagnetic kinetic instability known as the Weibel instability~\cite{weibel1959} can be triggered. The Weibel modes are typically purely growing modes that lead to exponential growth of magnetic fields by depleting the free energy stored in the pressure anisotropy. The wavevector of a transverse Weibel mode is in the direction of the smallest component of the pressure tensor, and the transverse magnetic fields (as well as their corresponding currents) are arranged in filamentary structures that are aligned perpendicularly to the wavevector.

The Weibel instability grows on a time scale proportional to the plasma frequency, $\omega_{{\rm p}s}$, for each species. 
The electron Weibel instability is thus much faster than that of ions.
In this subsection, we only consider the electron Weibel instability, triggered by the electron pressure anisotropy~$\Delta_e$.
The following discussion, and in particular the derived scaling laws, also applies to an electron-positron plasma (see the Supplementary Materials).

The linear theory of electromagnetic fluctuations in an initially unmagnetized bi-Maxwellian plasma indicates that, in the limit of weak anisotropy $\Delta_e \ll 1$, and considering only transverse modes, the electron Weibel growth rate, $\gamma_{\rm w}$, has a power-law dependence on the anisotropy, $\gamma_{\rm w} \simeq \Delta_e^{3/2} \omega_{{\rm p}e} v_{{\rm th} e}/c$, occurring at the most unstable mode with wavenumber $k_{\rm w} \simeq \sqrt{\Delta_e}/d_e$~\cite{weibel1959,davidson1972nonlinear}, where $d_e=c/\omega_{{\rm p}e}$ is the electron skin depth.
In our model, the evolving $f_s$ [Eq.~\eqref{eq:f_exact_uncharged_theory}] is not a bi-Maxwellian distribution, and therefore the dispersion relation of Weibel modes (and the scaling dependence of $\gamma_{\rm w}$ on $\Delta_e$) for this $f_s$ needs to be reexamined~\cite{silva2021weibel}. 
In the supplementary material, we show that for the short-time limit ($tv_{{\rm th}s}/L \ll 1$) and at a fixed position (e.g., $x=0$ with maximum shear), $f_s$ possesses the form of a multivariate normal distribution and becomes tri-Maxwellian in the coordinate system in which the axes are along the principal axes of the thermal pressure tensor.
In this case, the most unstable transverse mode has the same dispersion relation as that of the bi-Maxwellian distribution.
We also find that for a tri-Maxwellian plasma distribution and considering a general wavevector, the most unstable mode only has a transverse component (i.e., the longitudinal two-stream instability is subdominant), and the scaling dependence of its growth rate remains $\gamma_{\rm w} \simeq \Delta_e^{3/2} \omega_{{\rm p}e} v_{{\rm th} e}/c$.
This scaling holds for both an electron-positron plasma and an electron-ion plasma if the ions form a fixed, neutralizing background.

The short-time limit~($tv_{{\rm th}e/L}$) is relevant to most astrophysical environments where an asymptotically large separation between the kinetic ($1/\gamma_{\rm w}$) and the fluid ($L/v_{{\rm th}e}$) time scales exists. 
This is also the regime where the short-time approximation of the unmagnetized solution [Eqs.~\eqref{eq:fs_2nd_order}--\eqref{eq:Delta_2nd}] is valid.
We call this regime the asymptotic regime. 
The time scale separation $1/\gamma_{\rm w} \ll L/v_{{\rm th}e}$ is equivalent to the length scale separation $L/d_e \gg 1$ (with only a factor of order unity difference).

On the other hand, for systems lacking such a scale separation (such as those achievable in numerical simulations and laboratory laser experiments), at the moment when the Weibel magnetic fields are rapidly growing, $f_s$ already deviates significantly from a Maxwellian distribution and possesses a complex form (e.g., bottom panels in Fig.~\ref{fig:f0contour}).
In this case, a different Weibel dispersion relation is expected.
We assume that the dependence of the growth rate of the magnetic field, $\gamma_B$, on $\Delta_e$ remains a power law, and the power-law exponent is set to be a free parameter~$\alpha$:
\begin{equation}
\label{eq:gammab}
    \gamma_B  \equiv \frac{{\rm d} \ln B}{{\rm d}t}\sim \Delta_e^{\alpha}  \omega_{{\rm p}e} \frac{v_{{\rm th} e}}{c}.
\end{equation}
In the asymptotic regime, we expect $\alpha=3/2$.

During the linear stage of the Weibel instability, the magnetic field is not yet strong enough to affect the background accelerating plasma flow. 
The system should thus follow the unmagnetized solution [Eq.~\eqref{eq:f_exact_uncharged_theory}], based on which the evolution of $\Delta_e$ at arbitrary times does not have an explicit analytical expression. 
For simplicity, we assume a power-law scaling
\begin{equation}
    \Delta_e \sim \hat{a}_0 (t v_{{\rm th} e}/L)^{\kappa},
    \label{eq:delta_theory}
\end{equation}
where $\kappa=2$ in the asymptotic regime [Eq.\eqref{eq:Delta_2nd}].

As pressure anisotropy is continuously driven by the phase mixing, $\gamma_B$ also increases with time [Eq.~\eqref{eq:gammab}].
The time evolution of $\Delta_e$, and thus of $\gamma_B$, is a slow process on the fluid time scale $L/v_{{\rm th}e}$. 
The Weibel instability is a fast process on the kinetic time scale $1/\gamma_{\rm w}$.
If these two time scales are asymptotically separated, {\it viz.} $\gamma_B \gg \partial_t \Delta_e/\Delta_e \sim \partial_t \gamma_B/\gamma_B$, we can integrate Eq.~\eqref{eq:gammab} to obtain the evolution of the magnetic field. Assuming a constant mean thermal pressure of the system, the time evolution of $\beta_e^{-1}$ (representing magnetic energy) can then be written as
\begin{equation}
    \beta_e^{-1} \simeq \beta_{0}^{-1} \exp \left[\frac{2\hat{a}_0^\alpha}{\kappa \alpha+1} \left(\frac{t v_{{\rm th} e}}{L} \right)^{\kappa \alpha+1}\frac{L}{d_e} \right],
\label{eq:exponential_B_theory}
\end{equation}
where $\beta_{0}^{-1}$ is determined by the initial magnetic-field perturbation at~$k_{\rm w}$.

As the magnetic field keeps growing, it eventually becomes large enough to affect the trajectory of electrons significantly and, thus, the evolution of $\Delta_e$. 
At this point, Eq.~\eqref{eq:exponential_B_theory} is no longer valid.
The electron Weibel instability reaches the end of its linear stage and nonlinear effects start to play a role.
We denote this moment of time corresponding to the end of linear Weibel stage, normalized to $L/v_{{\rm th} e}$, as $\tau_{\rm lin}$.
Therefore, at $\tau_{\rm lin}$, the argument in the exponential function in Eq.~\eqref{eq:exponential_B_theory} is expected to reach order unity, resulting in the scaling
\begin{align}
    \tau_{\rm lin} \sim \left( \frac{L}{d_e} \right)^{-1/(\kappa \alpha+1)} \hat{a}_0^{-\alpha/(\kappa \alpha+1)}.
    \label{eq:tau_lin_scaling}
\end{align}
It follows that the electron pressure anisotropy $\Delta_e$ and the magnetic growth rate $\gamma_B$ at $\tau_{\rm lin}$ should satisfy
\begin{align}
    \label{eq:delta_lin_scaling}
    \Delta_e(\tau_{\rm lin}) &\sim \left( \frac{L}{d_e} \right)^{-\kappa/(\kappa \alpha+1)} \hat{a}_0^{1/(\kappa\alpha+1)},
    \\
    \frac{\gamma_B(\tau_{\rm lin})}{\omega_{{\rm p}e}} &\sim \left( \frac{L}{d_e} \right)^{-\kappa\alpha/(\kappa \alpha+1)} \hat{a}_0^{\alpha/(\kappa\alpha+1)}\frac{v_{{\rm th} e}}{c}.
    \label{eq:gammaB_lin_scaling}
\end{align}
In the asymptotic regime, we expect $\alpha=3/2$ and $\kappa=2$; the above scaling laws then become
\begin{align}
    \label{eq:tau_lin_scaling_largeLde}
    \tau_{\rm lin} &\sim \left( \frac{L}{d_e} \right)^{-1/4}  \hat{a}_0^{-3/8},\\
    \label{eq:delta_lin_scaling_largeLde}
    \Delta_e(\tau_{\rm lin}) &\sim \left( \frac{L}{d_e} \right)^{-1/2} \hat{a}_0^{1/4},\\
    \label{eq:gammaB_lin_scaling_largeLde}
    \frac{\gamma_B(\tau_{\rm lin})}{\omega_{{\rm p}e}} &\sim \left( \frac{L}{d_e} \right)^{-3/4} \hat{a}_0^{3/8} \, \frac{v_{{\rm th} e}}{c}.
\end{align}
The dependence of $\Delta_e(\tau_{\rm lin})$ on $L/d_e$ and $\hat{a}_0$ is essential for estimating the saturation level of Weibel magnetic fields, as we explain in the next subsection. 

\subsection{Saturation of Weibel instability.}
\label{sec:theory_saturation}

At $\tau_{\rm lin}$, $\Delta_e$ reaches its maximum value and the width of the forming Weibel filaments (the wave number of the Weibel modes) is determined by the value of $\Delta_e(\tau_{\rm lin})$. 
After $\tau_{\rm lin}$, the electron Weibel instability enters its nonlinear stage, during which we expect both its growth rate $\gamma_B$ and the anisotropy $\Delta_e$ to decrease rapidly as the free energy is converted into magnetic energy.
However, the length scale of the Weibel filaments should not change significantly in this nonlinear stage, instead remaining similar to that set by $\Delta_e(\tau_{\rm lin})$.
This is because the magnetic growth rate during the nonlinear stage is small compared to that of the linear stage. Although the wavenumber of the most unstable mode decreases together with~$\Delta_e$, we do not expect it to acquire much energy (an expectation confirmed by our simulation results; see Sec.~\ref{sec:fiducial}). Accordingly, the magnetic-energy-containing scale should remain similar to that achieved at the end of linear stage, when the magnetic growth rate is maximal and the Weibel filaments are fully formed. 
Other processes that can change the length scale of Weibel fields, such as the tilting of filaments due to the background shear flow and the coalescence of filaments, occur on time scales much longer than the inverse Weibel growth rate, and can thus be neglected before Weibel saturation occurs.

As the magnetic field becomes stronger, it affects the trajectories of particles and gradually magnetizes them. 
When electrons start to execute gyromotion with their Larmor radii, $\rho_e$, comparable to the length scale of the magnetic field, $k_{\rm w}^{-1}$, they are ``trapped'' in the Weibel filaments. This particle trapping condition, $k_{\rm w}\rho_e \sim 1$, is commonly believed to lead to the saturation of the electron Weibel instability~(e.g.,~\cite{davidson1972nonlinear,kato2005saturation}).

The dependence of the length scale of the Weibel magnetic field, $k^{-1}_{\rm w}(\tau_{\rm lin})$, on $\Delta_e(\tau_{\rm lin})$ is determined by the linear dispersion relation of the Weibel instability.
Alongside the power-law dependence of $\gamma_B$ on $\Delta_e$ [Eq.~\eqref{eq:gammab}], we also assume a power-law dependence of $k_{\rm w}$ on $\Delta_e$:
\begin{equation}
    k_{\rm w} \simeq \Delta_e^{\nu}/d_e,
    \label{eq:kw_delta}
\end{equation}
where we expect $\nu=1/2$ in the asymptotic regime (see~\cite{davidson1972nonlinear} and the Supplementary material).
It follows from Eq.~\eqref{eq:delta_lin_scaling} that the dependence of $k_{\rm w}d_e$ on $L/d_e$ and $\hat{a}_0$ satisfies
\begin{align} 
\label{eq:kwde_Lde}
     k_{\rm w}d_e \sim \left( \frac{L}{d_e} \right)^{-\kappa\nu/(\kappa \alpha+1)} \hat{a}_0^{\nu/(\kappa\alpha+1)} ;
\end{align}
in the asymptotic regime [using Eq.~\eqref{eq:delta_lin_scaling_largeLde}],
\begin{align} 
\label{eq:kwde_Lde_asym}
     k_{\rm w}d_e \sim \left( \frac{L}{d_e} \right)^{-1/4}  \hat{a}_0^{1/8}.
\end{align}
Thus, the dominant Weibel wavelength, $\lambda_{\rm w} = 2\pi/k_{\rm w}$, is a hybrid scale, intermediate between $L$ and $d_e$: 
\begin{equation}
\label{eq:lambdaw_Lde_asym}
    \lambda_{\rm w} \sim L^{1/4} d_e^{3/4} \hat{a}_0^{-1/8}.
\end{equation}

The average electron Larmor radius can be estimated as $\rho_e \simeq \beta_e^{1/2}d_e$. 
Combining this relation with Eq.~\eqref{eq:kw_delta}, the trapping condition, $k_{\rm w}\rho_e \sim 1$, provides the estimate of the value of $\beta_e^{-1}$ at saturation:
\begin{equation}
    \beta_{e,{\rm sat}}^{-1} \sim \Delta_e^{2\nu}(\tau_{\rm lin}).
    \label{eq:betasat_Delta}
\end{equation}
Combined with the relations in Eqs.~\eqref{eq:tau_lin_scaling}--\eqref{eq:gammaB_lin_scaling_largeLde}, we obtain the dependence of the saturated $\beta_e^{-1}$ on the system parameters:
\begin{align}
     \beta_{e,{\rm sat}}^{-1} \sim  \left( \frac{L}{d_e} \right)^{-\frac{2 \nu \kappa}{\kappa \alpha+1}} \hat{a}_0^{\frac{2\nu}{\kappa\alpha+1}}.
    \label{eq:betasat_Lde}
\end{align}
In the asymptotic regime ($\alpha=3/2$, $\kappa=2$, and $\nu=1/2$), this expression becomes
\begin{align}
     \beta_{e,{\rm sat}}^{-1} \sim  \left( \frac{L}{d_e} \right)^{-1/2} \hat{a}_0^{1/4}.
    \label{eq:betasat_Lde_largeLde}
\end{align}

Eqs.~\eqref{eq:kwde_Lde}--\eqref{eq:lambdaw_Lde_asym} and~\eqref{eq:betasat_Lde}--\eqref{eq:betasat_Lde_largeLde} provide the main deliverable of this study---the scaling dependence of the length scale [${\propto}(k_{\rm w}d_e)^{-1}$] and amplitude (${\propto}\beta_{e,{\rm sat}}^{-1}$) of the saturated seed magnetic fields on the two key dimensionless parameters: $L/d_e$ and $\hat{a}_0$.
The $\hat{a}_0$ is determined by the drive and related to the Mach number of the system [Eq.~\eqref{eq:Msat}].
Setting $L/d_e$ as a parameter allows us to test the predicted scalings [Eq.~\eqref{eq:gammab}--\eqref{eq:betasat_Lde_largeLde}] using numerical simulations with relatively small values of $L/d_e$, and then extrapolate to relevant astrophysical systems with asymptotically large $L/d_e$.
Note that another fundamental quantity in astrophysical environments --- the normalized temperature $\theta_s \equiv T_s/m_sc^2$ --- is not a critical parameter for this problem since we focus only on the sub-relativistic regime.
The Weibel magnetic energy and the thermal pressure are both proportional to $\theta_s$. Therefore, the saturated $\beta_e^{-1}$, reflecting the level of magnetization that can be achieved through the Weibel instability, is not a function of temperature (at fixed $\hat{a}_0$).

In this section, we have discussed in detail the response of an initially unmagnetized collisionless plasma to an externally driven, large-scale shear flow.
During the initial unmagnetized stage, an analytical solution of the plasma distribution function has been obtained, based on which we have calculated the time evolution of the Mach number $M_s$ and the pressure anisotropy $\Delta_s$ (Sec.~\ref{sec:theory_unmagnetized}).
The Weibel instability is triggered by the developed thermal anisotropy.
In the linear Weibel stage, we have used the unmagnetized solution of $f_s$ as the background equilibrium and performed linear theory to obtain the growth of magnetic energy, represented by $\beta_e^{-1}$ (Sec.~\ref{sec:theory_linear} and the Supplementary materials).
The length scale of the Weibel fields ($k_{\rm w}^{-1}$) determines their saturation level governed by the trapping condition; the dependence of $\beta_{e,{\rm sat}}^{-1}$ on the two key parameters ($L/d_e$ and $\hat{a}_0$) have thus been obtained (Sec.~\ref{sec:theory_saturation}). 

Our model is predictive for the scaling dependence of the dominant wavenumber [Eq.~\eqref{eq:kwde_Lde_asym}] and inverse beta [Eq.~\eqref{eq:betasat_Lde_largeLde}] for the saturated fields in the asymptotic regime: $k_{\rm w}d_e \sim (L/d_e)^{-1/4}\hat{a}_0^{1/8}$ and $\beta_{e,{\rm sat}}^{-1} \sim (L/d_e)^{-1/2}\hat{a}_0^{1/4}$.
In regimes lacking a large enough scale separation $L/d_e$, we have to set the exponents ($\alpha$, $\kappa$, and $\nu$) of certain power-law dependencies [Eqs.~\eqref{eq:gammab}--\eqref{eq:delta_theory} and \eqref{eq:kw_delta}] as undetermined parameters. 
Those exponents are to be determined by the first-principles numerical simulations discussed in Sec.~\ref{sec:numerical_results}. 
However, the derived scalings based on these undetermined exponents [Eqs.~\eqref{eq:tau_lin_scaling}--\eqref{eq:gammaB_lin_scaling}, \eqref{eq:kwde_Lde}, \eqref{eq:betasat_Delta}--\eqref{eq:betasat_Lde}] will be tested independently using the numerical results to validate the model.

\section{Simulation setup}
\label{sec:setup}
To test and calibrate the theory in Sec.~\ref{sec:theory}, we perform first-principles particle-in-cell (PIC) simulations using the code ZELTRON~\cite{Cerutti2013} of an initially unmagnetized plasma driven by an external shearing force.
Due to the high computational cost inherent to this problem, our simulations are performed using an electron-positron plasma ($s \in \{e,p\}$).
In the case that the external force causes the same body acceleration to both species, giving rise to a hydrodynamic flow, the evolution of an electron-positron system should be similar to an electron-proton plasma within the characteristic electron time scale (before the subsequent ion Weibel instability becomes active).
In the remainder of the paper, we drop the subscript $s$ and use $v_{\rm th}$ and $\omega_{\rm p}$ to represent the thermal velocity and plasma frequency for both electrons and positrons.
We set the initial temperatures to $\theta \equiv T/m_ec^2=1/16$ (so that the thermal motions of the particles are sub-relativistic). The thermal velocity is $v_{\rm th} \equiv \sqrt{T/m_e} = \sqrt{\theta} c$.
The system is initialized with uniform Maxwellian distributions and no electromagnetic fields, and is continuously driven by an external mechanical force $ \bb{F}_{\rm ext} = m_e \bb{a}$, where $\bb{a}=a_0 \sin{(2\pi x/L)} \hat{\bb{y}}$, as described in Sec.~\ref{sec:theory}.
We parameterize the acceleration amplitude $a_0=S_0 (\pi^2 \theta_e c^2/L)$, where $S_0$ is a parameter we control in the simulations and is related to the normalized forcing amplitude as $S_0=\hat{a}_0/\pi^2$.

The system is intrinsically multi-scale, containing the macroscopic, slow, fluid-scale dynamics driven by the external shear force; and the fast, kinetic-scale dynamics of plasma instabilities.
In order to explore both the slow and fast dynamics, we perform parameter scans on the two key parameters: $S_0$ and $L/d_e$.
Both 3D and 2D runs are performed with the same setup, with the 2D runs resolving only the $x$-$y$ plane (but including all three velocity components). 
The main purpose of the 2D runs is to achieve the largest values of $L/d_e$ that we can afford, and thus a better separation between the macro- and microscopic dynamics.
The dynamics in the unmagnetized stage (Sec.~\ref{sec:theory_unmagnetized}) is identical between 2D and 3D systems, and we expect their Weibel physics to be qualitatively similar---the scaling laws [Eqs.~\eqref{eq:gammab}-\eqref{eq:betasat_Lde_largeLde}] hold for both 2D and 3D cases with only a constant factor difference. 
On the other hand, the 2D runs do not capture possible dynamics in the $z$ direction such as the kink instability and the coalescence of Weibel filaments.
However, we will find (in Sec.~\ref{sec:numerical_results}) that those dynamics only affect the long-term evolution of Weibel filaments and do not change the main deliverable of this study: the scaling dependence of saturated Weibel seed fields on $L/d_e$ and~$S_0$.

We conduct scans in $S_0$ and $L/d_e$. For the scan in~$S_0$, which we vary across $S_0 \in \{0.1,0.2,0.3,0.4\}$, we perform one group of 3D runs with fixed $L/d_e=32$, and two groups of 2D runs with fixed $L/d_e=512$ and $L/d_e=1024$, respectively.
For the scan in $L/d_e$, we perform a group of 3D runs with fixed $S_0=0.2$ and varying $L/d_e \in \{32,48,64,96,128,192\}$, and a group of 2D runs with fixed $S_0=0.2$ and varying $L/d_e \in \{32,48,64,96,128,192,256,384,512,769,1024\}$.
For all simulations, the (initial) Debye length $\lambda_{{\rm D}e}=\Delta x$ where $\Delta x$ is the cell length, and $d_e=4 \Delta x$ (so that $d_e/\lambda_{{\rm D}e}=\sqrt{1/\theta_e}=4$). 
All 2D runs are performed using 256 particles per cell (PPC) (128 per species).
The 3D runs with fixed $S_0=0.2$ and varying $L/d_e$ are performed with 32 PPC, and those with fixed $L/d_e=32$ and varying $S_0$ have 256 PPC (for which the results are similar to those in runs with 32 PPC with all the other parameters kept identical).
All runs are evolved for more than one thermal crossing time to include both the micro- and macroscopic dynamics.

For the scan in $S_0$, the scale separation $L/d_e$ is fixed. We vary the amplitude of the forcing to the system and study how the kinetic physics responds to it.
For the scan in $L/d_e$, the system size $L$ is kept fixed and $d_e$ is varied by changing the plasma density. In other words, we drive the fluid-scale dynamics identically and study how the system's kinetic-scale response changes with scale separation.

\section{Numerical results}
\label{sec:numerical_results}

We first analyze in detail one single representative case: the 3D run with $L/d_e=128$ and $S_0=0.2$ (Sec.~\ref{sec:fiducial}).
The value of $L/d_e$ in this run is moderate so that we can both have a separation between the fluid-scale and Weibel dynamics and a long enough time interval to test our predictions for the unmagnetized stage.
We then use the scans in the input parameters $L/d_e$ and $S_0$ to test the scaling laws predicted by our model [Eqs.~\eqref{eq:tau_lin_scaling}--\eqref{eq:betasat_Lde_largeLde}] (Sec.~\ref{sec:sim_scalings}).

\subsection{Qualitative analysis of a fiducial case}
\label{sec:fiducial}
In this section we focus on the 3D run with $L/d_e=128$ and $S_0=0.2$. 

\begin{figure}
    \centering
    \includegraphics[width=0.45\textwidth]{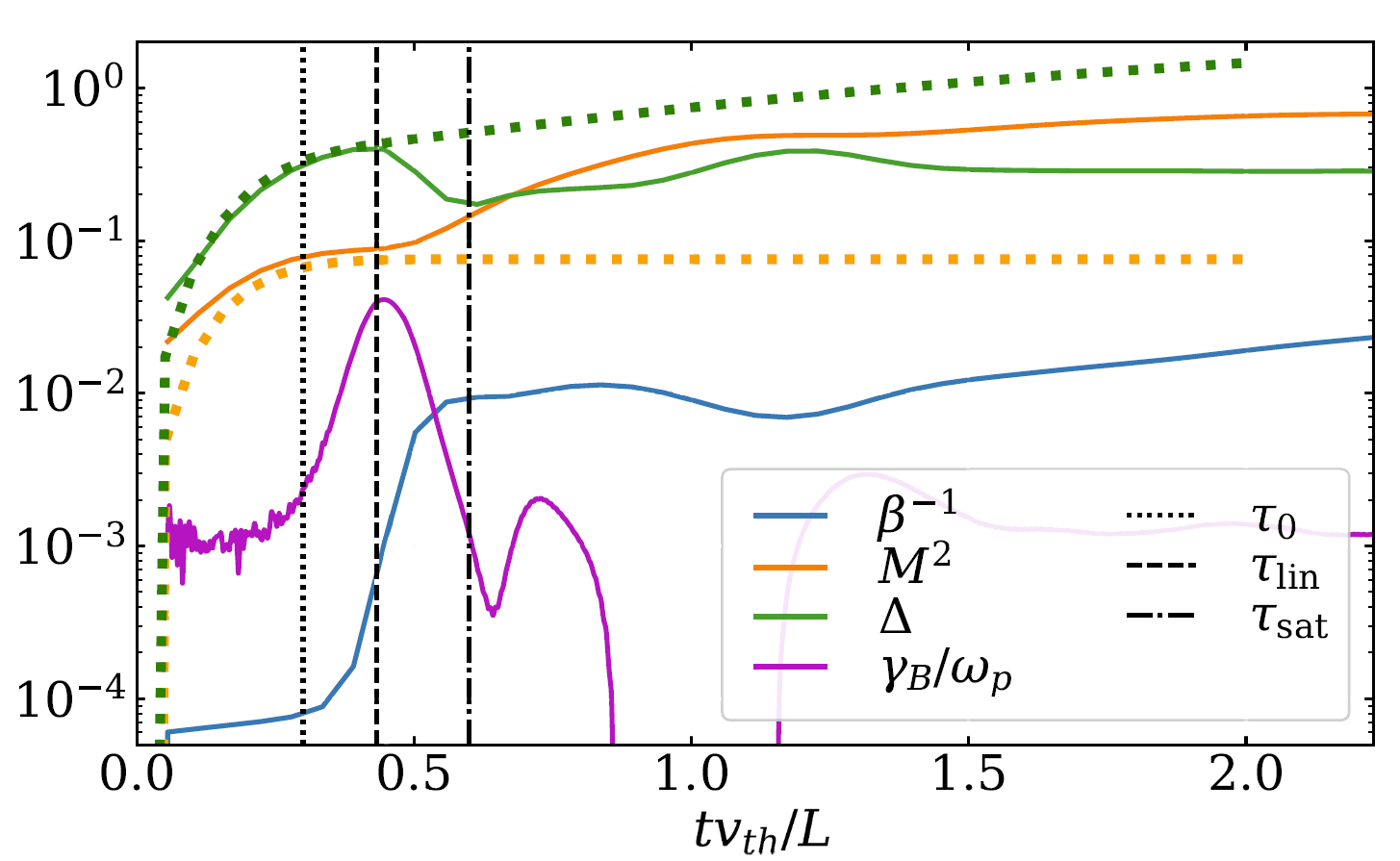}
    \caption{Time evolution of $M^2$, $\Delta$, $\beta^{-1}$, and $\gamma_B/\omega_{\rm p}$ from the run with $L/d_e=128$ and $S_0=0.2$. Dotted lines show the analytical results calculated with the unmagnetized solution Eq.~\eqref{eq:f_exact_uncharged_theory}. }
    \label{fig:512_M_delta_beta}
\end{figure}

The measured Mach number $M$, pressure anisotropy~$\Delta$, and plasma beta $\beta$ are identical between the two species and are therefore written without a species subscript. 
Fig.~\ref{fig:512_M_delta_beta} compares the time evolution of $M^2$, $\Delta$, $\beta^{-1}$, and $\gamma_B/\omega_{\rm p}$.
The evolution of the system can be divided into four stages: the initial unmagnetized stage, the linear Weibel stage ($tv_{\rm th}/L \lesssim \tau_{\rm lin}$), the nonlinear Weibel stage ($\tau_{\rm lin} < tv_{\rm th}/L \leq \tau_{\rm sat}$), and a prolonged stage after the saturation of Weibel instability ($tv_{\rm th}/L > \tau_{\rm sat}$).
We describe each distinct stage qualitatively to reveal the physical picture of the whole process.  

\subsubsection{Growth of pressure anisotropy during unmagnetized stage.}

In the initial, unmagnetized, stage, the measured evolution of $M(t)$ and $\Delta(t)$ (shown in Fig.~\ref{fig:512_M_delta_beta}) agrees reasonably well with the analytical prediction obtained by numerically integrating the exact solution of $f_e$ in Eq.~\eqref{eq:f_exact_uncharged_theory} (shown by the dotted curves). 
The slight departure from the prediction at very early times is due to numerical noise from the finite number of particles. 
The development of thermal pressure anisotropy $\Delta$ is due to the phase mixing of particles and is a purely kinetic feature of the collisionless plasma.
In Sec.~\ref{sec:theory_unmagnetized}, we predicted that, in an unmagnetized plasma, the bulk flow velocity, and thus $M$, should saturate due to the developed effective viscous force that balances the external forcing.
This is indeed observed in the numerical results as the $M^2$ curve reaches a plateau after $\tau_0$ ($\approx 0.25$; dotted vertical line). 

\subsubsection{Growth of magnetic fields during linear Weibel stage.}
\label{sec:sim_linear}

\begin{figure*}[t!]
    \includegraphics[width=1.0\textwidth]{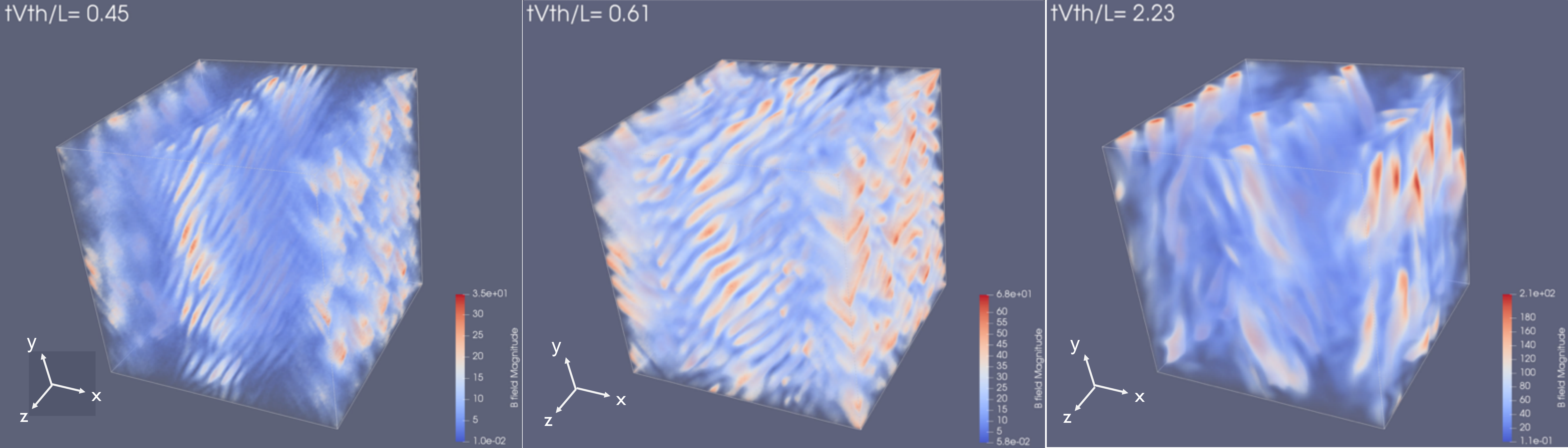}
    \caption{Visualization of magnetic-field amplitude at $\tau_{\rm lin}$ (left), $\tau_{\rm sat}$ (middle), and the end (right) of the the fiducial run with $L/d_e=128$.}
    \label{fig:visual_Bfield}
\end{figure*}
\begin{figure*}[t!]
    \centering
    \includegraphics[width=1.0\textwidth]{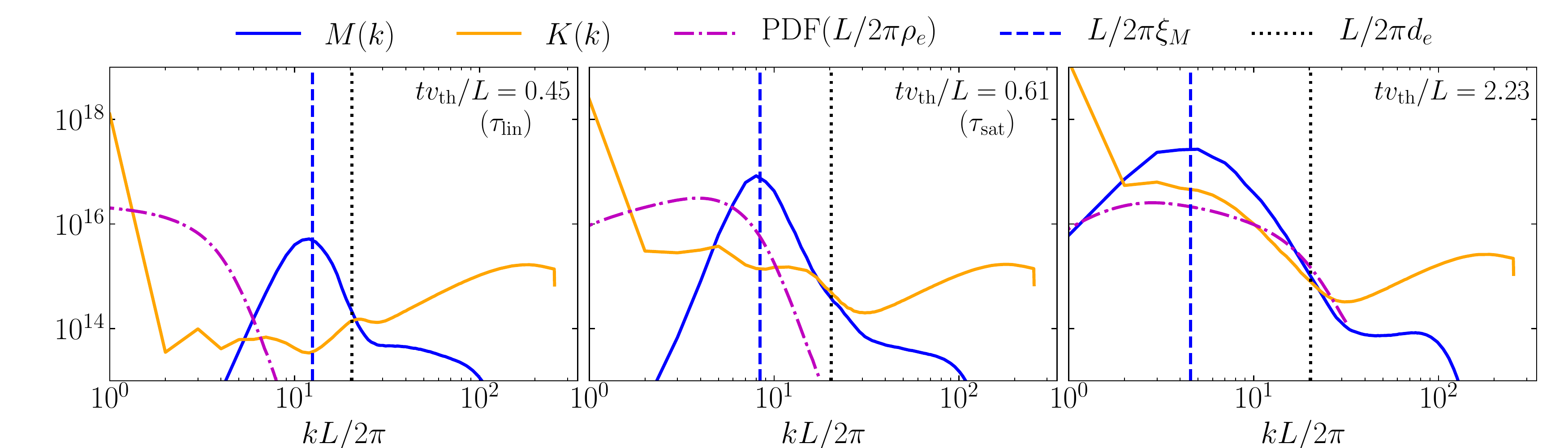}
    \caption{Magnetic (blue) and kinetic (orange) energy spectra at $\tau_{\rm lin}$ (left), $\tau_{\rm sat}$ (middle), and the end (right) of the 3D run with $L/d_e=128$. The electron skin depth $d_e$ (dotted vertical lines), magnetic energy integral scale $\xi_M$ (dashed vertical lines), and the PDF of Larmor radius $\rho_e$ (magenta dashed curve) are shown for reference.}
    \label{fig:512_mag_spec_tau1tau2}
\end{figure*}
\begin{figure}[h]
    \includegraphics[width=0.45\textwidth]{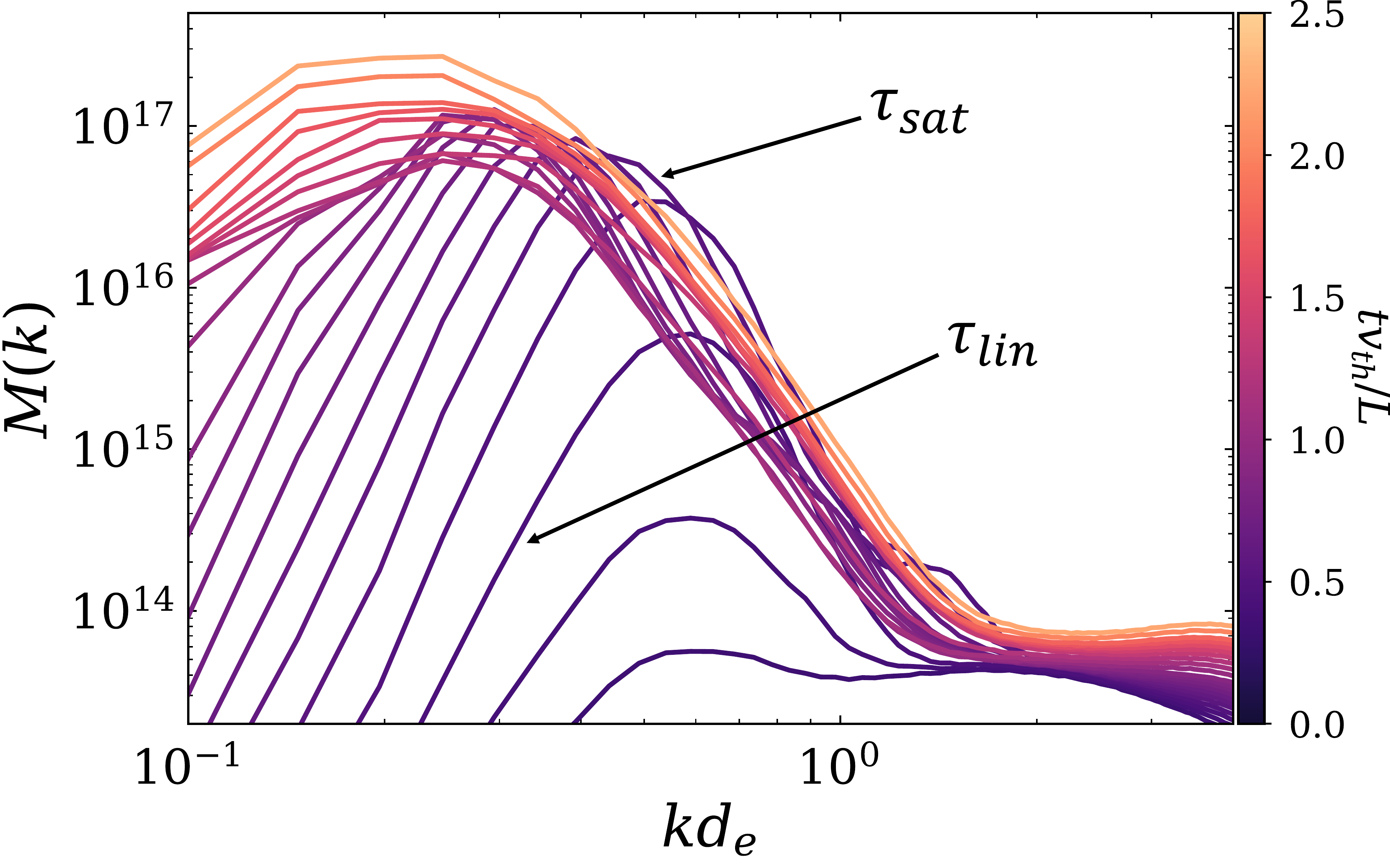}
    \caption{Time evolution of magnetic energy spectrum for the 3D run with $L/d_e=128$.}
    \label{fig:512_magspec_evo}
\end{figure}

With the development of pressure anisotropy ($\Delta$), the magnetic fields, and thus $\beta^{-1}$, start to grow exponentially as a result of the Weibel instability.
Fig.~\ref{fig:512_M_delta_beta} shows that in this linear Weibel stage, the measured magnetic growth rate, $\gamma_B \equiv \rmd\ln B/\rmd t$, also increases with time, suggesting a super-exponential growth of magnetic fields.
Magnetic fields with clear Weibel-type filamentary structures emerge on ${\sim}d_e$ scales from the initial random noise (Fig.~\ref{fig:visual_Bfield}, left panel). 

We identify a noteworthy moment of time, $\tau_{\rm lin}$ ($\approx 0.45$; vertical dashed line in Fig.~\ref{fig:512_M_delta_beta}), when the system's dynamics change in a qualitative manner. This is the time at which both $\Delta$ and $\gamma_B$ reach their maxima and then start a sharp downturn, while $M^2$ deviates from the plateau and starts to increase again. 
The $\beta^{-1}$ continues its exponential growth but at a relatively smaller rate. 
At $\tau_{\rm lin}$, both $\Delta$ and $M^2$ begin to depart from the (unmagnetized) analytical solution. 
These observations suggest that $\tau_{\rm lin}$ is the moment at which the Weibel magnetic fields have reached a magnitude sufficient to affect the dynamics of the plasma---i.e., nonlinear effects become important.

Power spectra of fluctuations [integrated isotropically in wavenumber ($\bb{k}$) space] at $\tau_{\rm lin}$ are shown in the left panel of Fig.~\ref{fig:512_mag_spec_tau1tau2}.
The power spectrum of the bulk flow, $K(k)$ is concentrated at the system scale where the flow is driven. 
In contrast, the power spectrum of the magnetic field, $M(k)$, peaks at ${\sim}d_e$ scale, consistent with the structure of the magnetic filaments shown in the left panel of Fig.~\ref{fig:visual_Bfield}. 
We define the magnetic-energy containing scale (shown by the blue vertical dashed line in Fig.~\ref{fig:512_mag_spec_tau1tau2}) as $\xi_M \equiv \int\rmd k \, k^{-1} M(k)/\int\rmd k\, M(k)$, which is expected to relate to the wavenumber of the most unstable Weibel modes as $k_{\rm w}\xi_M \sim 1$. 
The dashed magenta curve shows the probability density function (PDF) of electron Larmor radius, $\rho_e\equiv m_e v_{\rm th}/(eB)$, where $v_{\rm th}$ and $B$ correspond to the local temperature and magnetic field on the numerical grid.
The local temperature is calculated by averaging the three diagonal elements of the local thermal pressure tensor defined in Eq.~\eqref{eq:define_Ps}.
At $\tau_{\rm lin}$, the plasma remains unmagnetized as the Larmor radii of the majority of particles are generally of order~$L/2\pi$, substantially larger than the scale of the magnetic field~$\xi_M$.

The evolution of the magnetic spectrum is shown in Fig.~\ref{fig:512_magspec_evo}.
In the linear Weibel stage, the amplitude of the spectrum increases rapidly while its peak has a slight shift to the larger wavenumbers, consistent with the increase of $\Delta$ during this stage.

\subsubsection{Saturation of Weibel instability during nonlinear stage.}
\label{sec:sim_single_saturation}

After $\tau_{\rm lin}$, the Weibel instability enters its nonlinear stage, in which the Weibel magnetic fields are strong enough to affect the particle trajectories and affect the overall plasma dynamics.
The pressure anisotropy $\Delta$ decreases as its free energy is depleted by the Weibel instability, resulting in a drop in $\gamma_B$ (Fig.~\ref{fig:512_M_delta_beta}).
The nonlinear Weibel instability saturates at the moment of time that we denote as~$\tau_{\rm sat}$ ($\approx 0.61$). 
At this time, $\beta^{-1}$ saturates (at the value that we call $\beta^{-1}_{\rm sat}$), $\gamma_B$ drops to a minuscule value, and $\Delta$ reaches its local minimum because the depletion of free energy in pressure anisotropy stops as Weibel instability saturates.
We use the local minimum of $\Delta$ in simulations to identify~$\tau_{\rm sat}$.

The configuration of magnetic fields at $\tau_{\rm sat}$ is shown in the middle panel in Fig.~\ref{fig:visual_Bfield}.
The filamentary structures become more prominent with stronger field amplitudes and the filaments become progressively tilted due to the large-scale shear flow along the $y$-axis.
The spectra during this stage are shown in the middle panel in Fig.~\ref{fig:512_mag_spec_tau1tau2}.
The characteristic scale of the magnetic field $\xi_M$ has increased by about $50\%$ as a combined effect of the decreasing unstable wavenumber due to the decreasing $\Delta$ and the tilting of filaments. 
The relatively modest increase in $\xi_M$ justifies our assumption in Sec.~\ref{sec:theory_saturation} that the length scale of magnetic fields at $\tau_{\rm sat}$ is similar to that at~$\tau_{\rm lin}$.
In contrast, the magnetic energy has increased by more than an order of magnitude between $\tau_{\rm lin}$ and $\tau_{\rm sat}$. 
This rapid growth of the magnetic field's amplitude and the slow change of its characteristic length scale during the nonlinear Weibel stage are illustrated in Fig.~\ref{fig:512_magspec_evo}. 
Next, we observe that some bulk kinetic energy develops near the scale of the magnetic field (kinetic scales), corresponding to bulk motions of the filaments. However, the energy of these motions is subdominant to the magnetic energy at those scales. 
No strong turbulent cascade develops, and the bulk flow remains concentrated at the macroscopic system scale.

\begin{figure}
    \centering
    \includegraphics[width=0.45\textwidth]{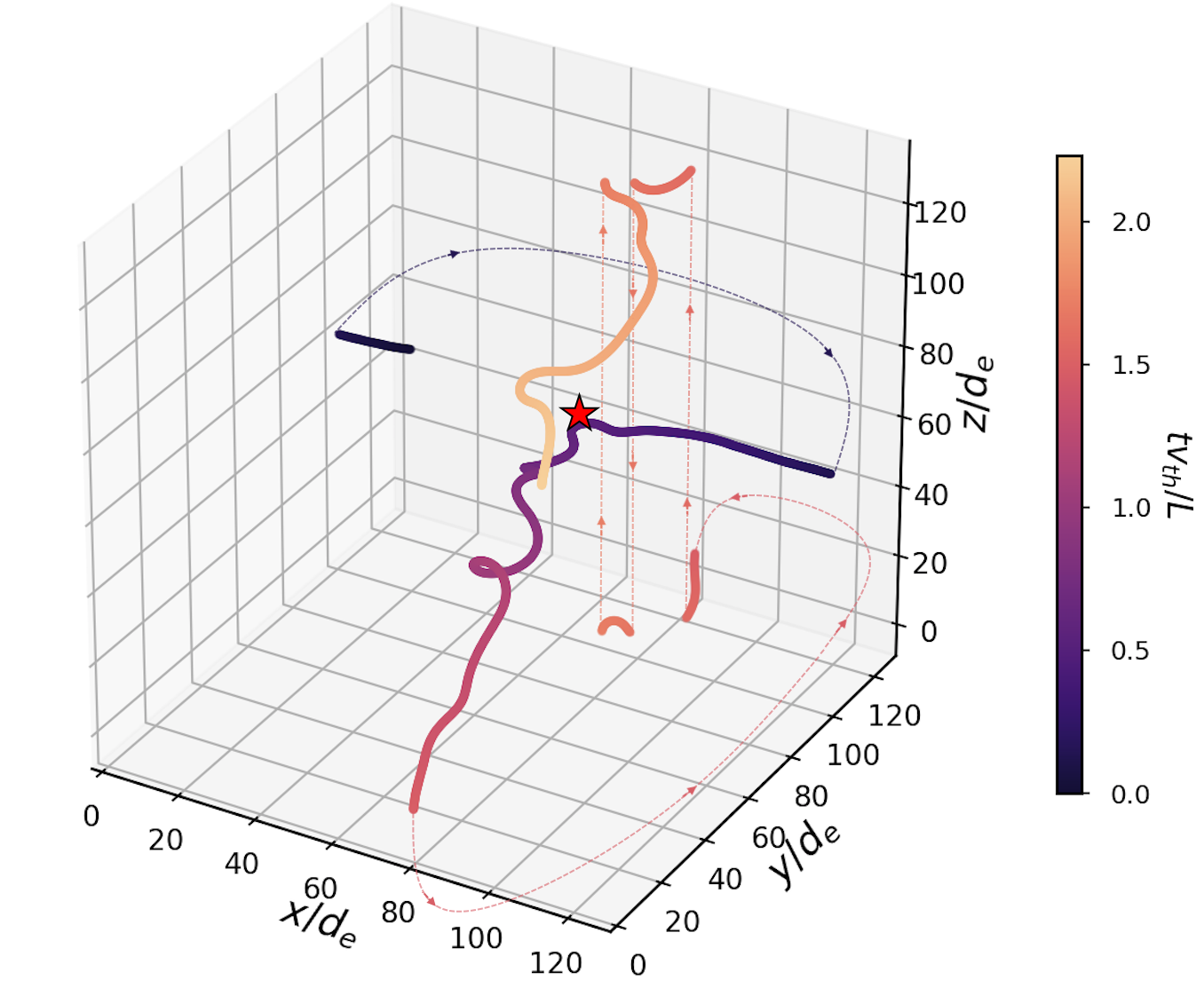}
    \caption{Typical trajectory of a particle from the run with $S_0=0.2$ and $L/d_e=128$. The red star indicates the particle's position at $\tau_{\rm sat}$. The dotted lines indicate how the particle transits the periodic box.}
    \label{fig:trajectory}
\end{figure}

The Larmor radii of a significant fraction of particles at $\tau_{\rm sat}$ become smaller than the scale of magnetic fields~$\xi_M$, meaning that those particles are magnetized by the Weibel magnetic fields. 
The magnetization of the plasma is also reflected in the trajectories of particles. 
Fig.~\ref{fig:trajectory} shows a representative trajectory of an arbitrarily chosen particle.
The particle initially streams freely along the $x$ and $z$ directions while being pushed by the external force in the $\pm y$ direction.
After $t v_{\rm th}/L=\tau_{\rm sat}$, the particle is trapped in the magnetic filaments in the $y$ and $z$-directions, while its transport in the $x$-direction is suppressed. 
This particle trapping leads to the suppression of the $y$-momentum transport in the $x$-direction, and hence to a dramatic reduction in the effective viscosity (see Sec.~\ref{sec:theory_unmagnetized}). As a result, the force balance in the $y$-direction is broken and the bulk flow starts to accelerate again driven by the external force, so that the Mach number $M$ starts to increase again rapidly around this time.
The above evidence from spectra and particle trajectories suggests that the saturation of the Weibel instability that we observe is caused by the trapping of particles, i.e., it occurs when the condition $k_{\rm w} \rho_e \sim 1$ is met --- a standard criterion widely considered by previous studies~(e.g.,~\cite{davidson1972nonlinear,kato2005saturation}).

\subsubsection{Long-term evolution of Weibel magnetic fields.}

After the saturation of the Weibel instability, on time scales of order $L/v_{\rm th}$, $\beta^{-1}$ fluctuates around $\beta_{\rm sat}^{-1}$, $M$ keeps increasing, and $\Delta$ starts to increase again due to the external forcing (Fig.~\ref{fig:512_M_delta_beta}). 
The saturated magnetic filaments are tilted and stretched by the shear flow until they become aligned in the direction of the shear flow (along the $y$-axis), as shown in the right panel of Fig.~\ref{fig:visual_Bfield}.
Alongside their interaction with the shear flow, the magnetic filaments also undergo a prolonged stage of coalescence with each other~\cite{zhou2019magnetic,zhou2020multi,zhou2021statistical}, during which the coherence length of magnetic fields increases. 
This can be seen from the shift of the power spectrum of magnetic fields to smaller wavenumbers (shown in Fig.~\ref{fig:512_magspec_evo}).
From $tv_{\rm th}/L=\tau_{\rm sat}$ to the end of simulation (with a time interval of about $1.6L/v_{\rm th}$), the energy-containing scale of the magnetic field grows to approach the system scale, at which magnetic energy accumulates.
Fig.~\ref{fig:512_mag_spec_tau1tau2}, right panel shows the spectra at $tv_{\rm th}/L=2.23$.
At this late time, the bulk kinetic energy increases at the scale of filaments but remains subdominant except at the system scale. 
Because of the combination of the slight increase of magnetic energy due to the transient inductive amplification by the shear flow, and the growth of the magnetic-field length scale $\xi_M$ through filament coalescence, more particles become magnetized (shown by the PDF of Larmor radii compared to the scale of $\xi_M$).

\begin{figure}
    \centering
    \includegraphics[width=0.45\textwidth]{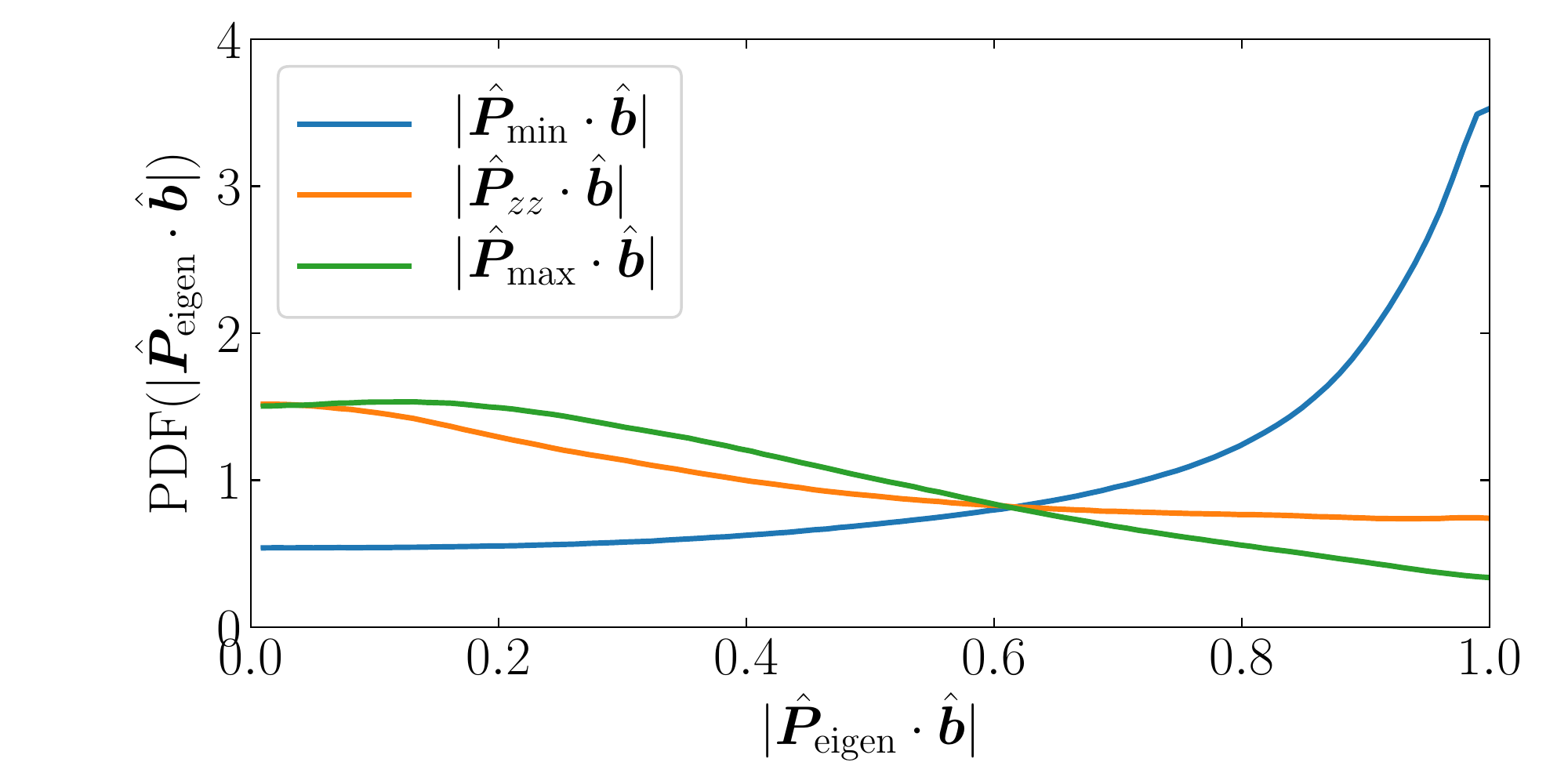}
    \caption{PDF of the alignment of magnetic-field direction $\hat{\bb{b}}$ with the thermal-pressure eigenvectors $\hat{\bb{P}}_{\rm eigen} \in \{\hat{\bb{P}}_{\rm min}, \hat{\bb{P}}_{zz},\hat{\bb{P}}_{\rm max} \}$, measured at $tv_{\rm th}/L=2.23$ of the 3D run with $L/d_e=128$.}
    \label{fig:P_B_alignment}
\end{figure}

The increasing magnetization can also be quantified by the alignment between the eigenvectors of the pressure tensor $\msb{P}$ and the local magnetic-field unit vector $\hat{\bb{b}}$.
We denote $\hat{\bb{P}}_{\rm min}$, $\hat{\bb{P}}_{zz}$, and $\hat{\bb{P}}_{\rm max}$ as the three eigenvectors corresponding to the three eigenvalues $P_{\rm min}<P_{zz}<P_{\rm max}$ of $\msb{P}$.
Fig.~\ref{fig:P_B_alignment} shows the probability distribution functions (PDFs) of the alignments $|\hat{\bb{P}}_{\rm eigen} \bcdot \hat{\bb{b}}|$, where $\hat{\bb{P}}_{\rm eigen} \in \{\hat{\bb{P}}_{\rm min}, \hat{\bb{P}}_{zz},\hat{\bb{P}}_{\rm max} \}$, at the end of the simulation ($tv_{\rm th}/L=2.23$). 
The magnetic field is primary aligned with $\hat{\bb{P}}_{\rm min}$, while the PDFs of its alignment with the other two directions are very broad and similar to each other.
These statistics result from the magnetization of the particles, manifested via  the approximate conservation of the first adiabatic invariant $\mu \equiv P_{\perp B}/nB$, where $P_{\perp B}$ is the thermal pressure perpendicular to the magnetic field. 
As $B$ increases, the conservation of $\mu$ leads to a biased increase of $P_{\perp B}$, and so the direction of the smallest pressure should correspond to the magnetic-field direction. 
This is displayed by the measured large $|\hat{\bb{P}}_{\rm min} \bcdot \hat{\bb{b}}|$.
Magnetized plasmas are approximately gyrotropic perpendicular to the magnetic field, consistent with the similar statistics of $|\hat{\bb{P}}_{zz} \bcdot \hat{\bb{b}}|$ and $|\hat{\bb{P}}_{\rm max} \bcdot \hat{\bb{b}}|$.
The magnetization of a significant fraction of the plasma particles is crucial for the coalescence of seed-field filaments where magnetic reconnection is essential~\cite{zhou2019magnetic,zhou2020multi,zhou2021statistical}, and for the further amplification of the seed fields by the turbulent dynamo.  
\\

By analyzing this single representative case, we have found that the system's evolution agrees qualitatively with our model described in Sec.~\ref{sec:theory}. 
We proceed to test the scaling laws predicted by our model using more quantitative measurements from the numerical simulations.

\begin{figure*}
    \centering
    \includegraphics[width=1.0\textwidth]{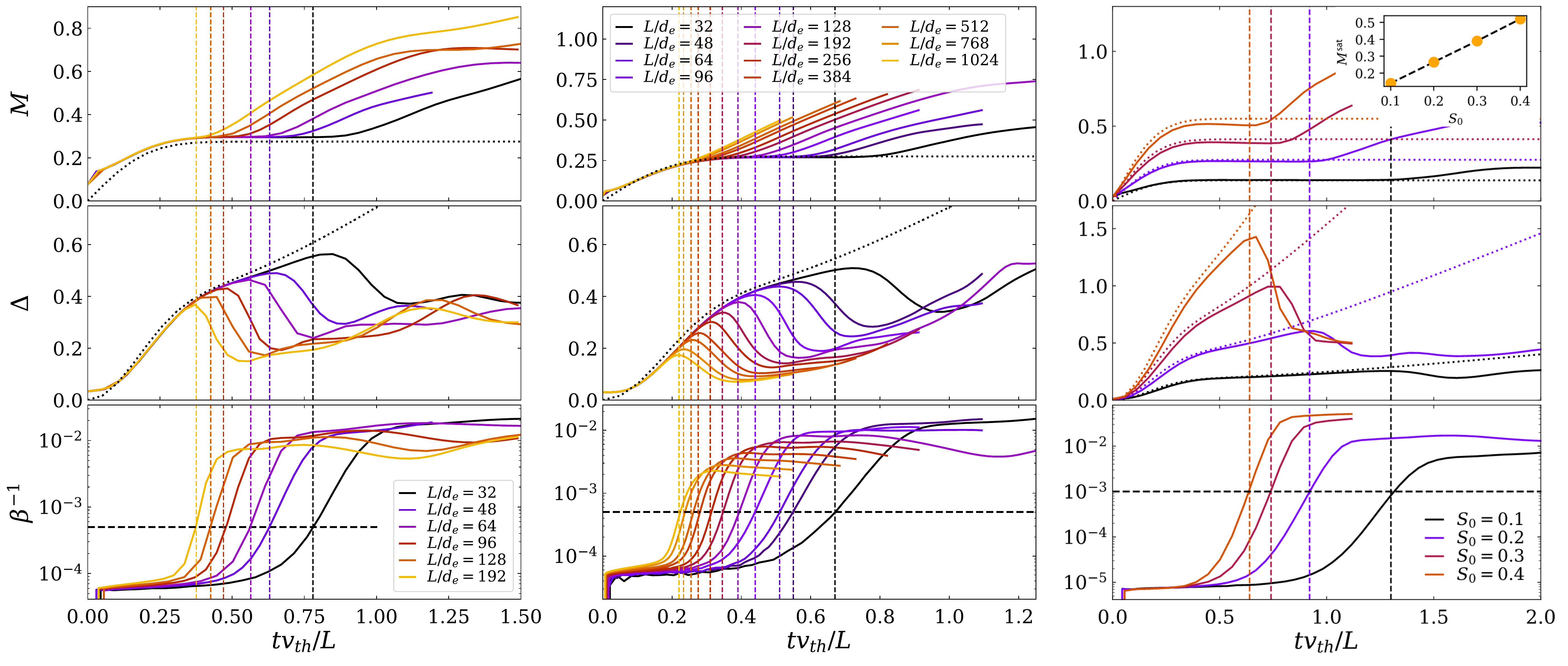}
    \caption{Time evolution of $M$ (top row), $\Delta$ (middle row), and $\beta^{-1}$ (bottom row). Left: 3D runs with varying $L/d_e$ and fixed $S_0=0.2$. Middle: 2D runs with varying $L/d_e$ and fixed $S_0=0.2$. Right: 3D runs with varying $S_0$ and fixed $L/d_e=32$. Vertical dashed lines indicate $tv_{\rm th}/L=\tau_{\rm lin}$ for corresponding runs. Horizontal dashed lines in the bottom panels of each column indicate the values of $\beta^{-1}$ at $\tau_{\rm lin}$. The dotted lines in the top and middle panels are the analytical solutions for $M$ and $\Delta$, respectively. The inset figure in the top-right panel shows the values of $M$ at the plateau versus $S_0$.}
    \label{fig:scanLdeS0}
\end{figure*}

\subsection{Quantitative scalings from parameter scans.}
\label{sec:sim_scalings}

We now focus on the parameter scans (in $S_0$ and $L/d_e$), analyzing the scaling laws of key quantities ($\Delta$, $\beta^{-1}$, and~$\gamma_B$) at critical moments of time ($\tau_{\rm lin}$ and $\tau_{\rm sat}$) and comparing our numerical results with the predictions derived in Sec.~\ref{sec:theory} [Eqs.~\eqref{eq:tau_lin_scaling}--\eqref{eq:gammaB_lin_scaling} and \eqref{eq:betasat_Delta}--\eqref{eq:betasat_Lde}].

The time evolution of $M$, $\Delta$, and $\beta^{-1}$ for these two parameter scans is shown in Fig.~\ref{fig:scanLdeS0}.
For runs performed at fixed $S_0$, during the unmagnetized and linear Weibel stages for each run, the evolution of macroscopic quantities ($M$ and $\Delta$) is identical (left and middle columns in Fig.~\ref{fig:scanLdeS0}).
For runs with varying $S_0$ (right column in Fig.~\ref{fig:scanLdeS0}), $M(t)$ and $\Delta(t)$ evolve differently, following Eq.~\eqref{eq:main_vlasov_theory}.
Simulations with different $L/d_e$ and $S_0$ enter the exponential magnetic-field growth stage at different moments of time. 
Even for systems sharing the same background evolution of $M(t)$ and $\Delta(t)$, their increase of $\beta^{-1}$ differs (left and middle column).
Systems with larger $L/d_e$ have a shorter kinetic time scale $\omega^{-1}_{\rm p}=d_e/c$ (relative to the macroscopic time scale $L/v_{\rm th}$) and thus a faster increase of $\beta^{-1}$ given that the growth rate of the Weibel instability $\gamma_B \propto \omega_{\rm p}$. 
Before entering the nonlinear Weibel stage, the magnetic-field strength is not yet significant enough to affect the macroscopic background evolution and, therefore, $M$ and $\Delta$ have not deviated from the unmagnetized solution (dotted lines).

In Sec.~\ref{sec:theory_unmagnetized}, we predict that, in an unmagnetized plasma, the bulk flow velocity, and thus $M$, should reach a saturation stage due to the developed effective viscous force that balances the external forcing.
In our numerical results, this feature is indeed observed for runs with $L/d_e \lesssim 200$. 
The force balance condition [Eq.~\eqref{eq:forcebalance}] provides an estimate of the plateau level $M^{\rm sat} \propto S_0$ [Eq.~\eqref{eq:Msat}]; this scaling is confirmed by the numerical results shown in the inset figure in the right column of Fig.~\ref{fig:scanLdeS0}.
For runs with $L/d_e \gtrsim 200$, the plateau of $M$ does not have enough time to develop because the overall dynamics is changed by the Weibel magnetic field before the force balance is reached.

In our simulations with fixed $S_0=0.2$, two regimes exist, depending on the scale separation~$L/d_e$.
For $L/d_e \lesssim 200$, the linear Weibel stage that occurs around $\tau_{\rm lin}$ is reached after $\tau_0$, the moment when the unmagnetized plasma reaches a steady-state flow and $M$ reaches the plateau. We call this the post-plateau regime.
For $L/d_e \gtrsim 200$, $\tau_{\rm lin}$ is reached before $\tau_0$.
Weibel fields grow shortly after the system is driven and change the overall dynamics before the steady-state flow could occur. We call this the pre-plateau regime. 
We denote by $(L/d_e)_{\rm cr}$ the critical scale separation where the transition between the pre- and post-plateau regimes occurs.  
Near this transition, the Weibel fields grow rapidly while the flow approaches the steady state, i.e., $\tau_0 \approx \tau_{\rm lin}$.
Combined with the estimation of these two times: $\tau_0 \sim 1/2\pi$ (see Sec.~\ref{sec:theory_unmagnetized}) and $\tau_{\rm lin} \sim (L/d_e)^{-1/(\kappa\alpha+1)}S_0^{-\alpha/(\kappa\alpha+1)}$~[Eq.~\eqref{eq:tau_lin_scaling}], we obtain the dependence of this critical scale separation on the drive of the system: $(L/d_e)_{\rm cr} \propto S_0^{-\alpha}$.

Most of our 3D simulations are in the post-plateau regime, with the largest ones ($L/d_e=128,192$) marginally entering the pre-plateau regime, while our 2D runs, where much larger values of $L/d_e$ can be afforded, allow us to explore the pre-plateau regime. The pre-plateau regime is closer to the asymptotic regime, which is relevant to astrophysical systems where $L/d_e$ is typically an asymptotically large number.
In the following subsections, we discuss the scaling laws measured during the the linear stage and saturation of the Weibel instability for both the pre- and post-plateau regimes.

\subsubsection{Scaling laws at the end of linear Weibel stage.}
\label{sec:sim_sclaing_linear}

\begin{figure}[h!]
    \centering
    \includegraphics[width=0.45\textwidth]{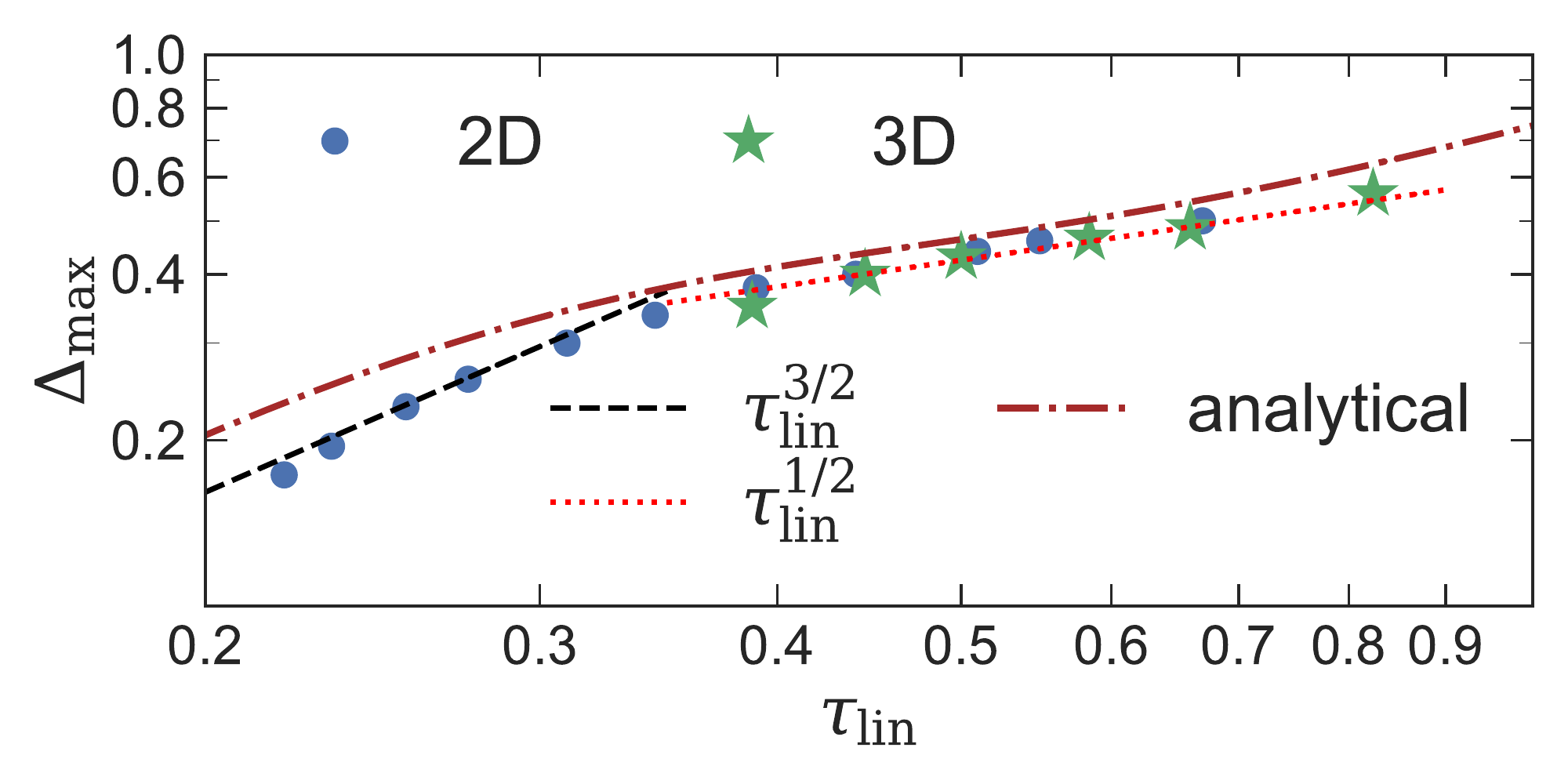}
    \caption{Results of $\Delta_{\rm max}$ versus $\tau_{\rm lin}$ from 2D and 3D runs with varying $L/d_e$ and fixed $S_0=0.2$. The dash-dotted curve shows the pressure anisotropy $\Delta$ as a function of time calculated from the analytical solution Eq.~\eqref{eq:f_exact_uncharged_theory}. Red-dotted and black-dashed lines show power-law fits to the post-plateau and pre-plateau regimes, respectively.}
    \label{fig:detla_taulin}
\end{figure}

\begin{figure}[h!]
    \centering
    \includegraphics[width=0.45\textwidth]{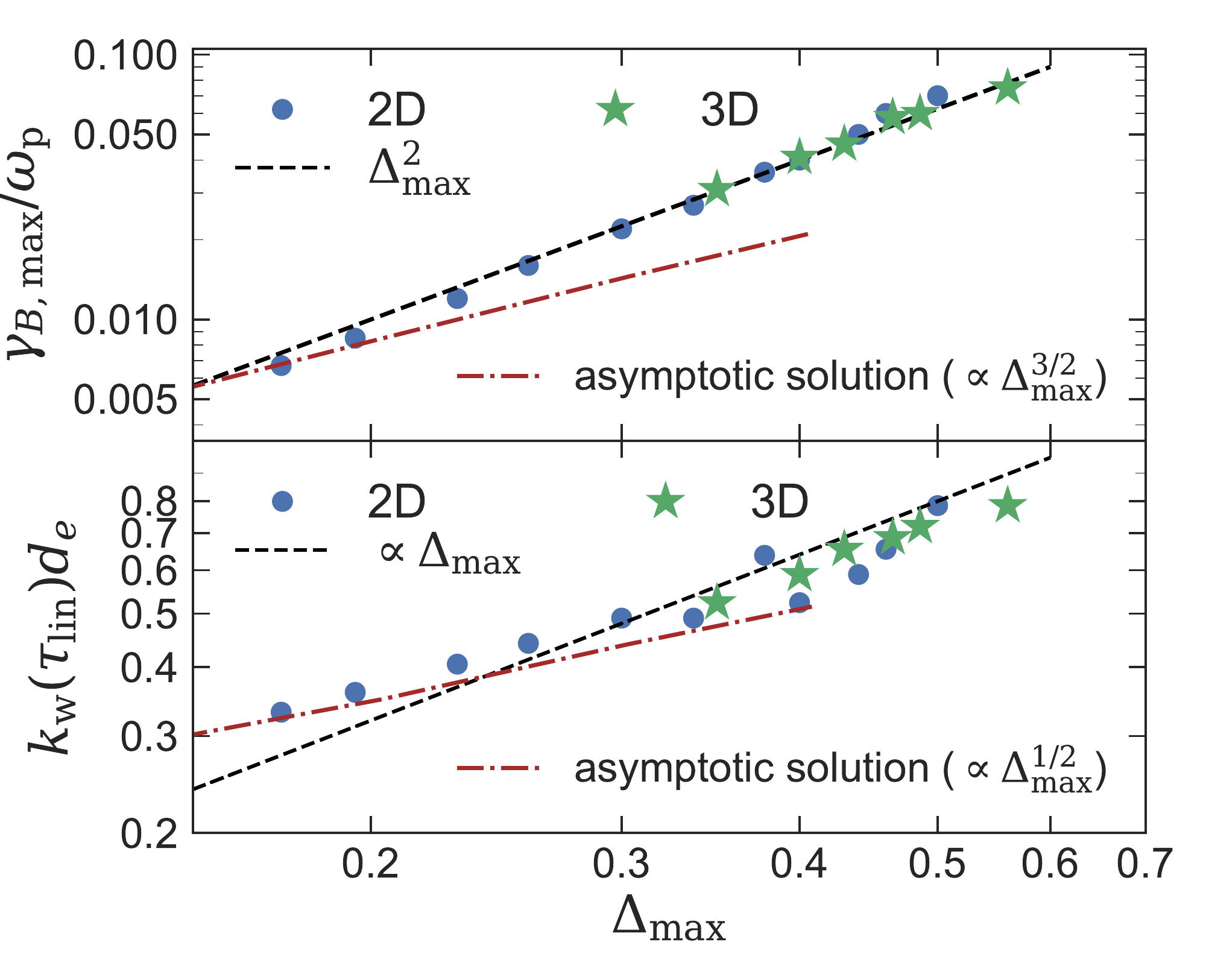}
    \caption{Weibel growth rate and wavenumber from 2D and 3D runs with varying $L/d_e$ and fixed $S_0=0.2$. Top: $\gamma_{B,{\rm max}}/\omega_{\rm p}$ versus $\Delta_{\rm max}$. The dashed line shows the $\sim \Delta_{\rm max}^2$ fit.  Bottom: $k_{\rm w}(\tau_{\rm lin})d_e$ versus $\Delta_{\rm max}$. The black dashed line shows a reference linear scaling.
    The brown dash-dotted lines show the asymptotic solution of the linear growth rate of the most unstable Weibel mode (top) and its corresponding wavenumber (bottom) as a function of pressure anisotropy. The values of measured growth rate and wavenumber from the two runs with the largest $L/d_e$ (the two left-most data points) agree with the asymptotic solution.}
    \label{fig:weibel_gamma_kw}
\end{figure}

In the linear Weibel stage, the plasma is unmagnetized and $\Delta$ increases due to the external forcing until reaching its maximum value $\Delta_{\rm max}$ at $\tau_{\rm lin}$, whereupon the effects of magnetic fields become important.
For runs with varying $L/d_e$, and thus varying $\tau_{\rm lin}$, the measured $\Delta_{\rm max}$ as a function of $\tau_{\rm lin}$ follows the time evolution of $\Delta$ calculated with the unmagnetized analytical solution Eq.\eqref{eq:f_exact_uncharged_theory}, as is shown in Fig.~\ref{fig:detla_taulin}.
The time evolution of $\Delta$, and thus the dependence of $\Delta_{\rm max}$ on $\tau_{\rm lin}$, can be approximated with power-law expressions within certain ranges of time: $\Delta \simeq \hat{a}_0 (tv_{\rm th}/L)^\kappa$ with $\hat{a}_0 \propto S_0$ [Eq.~\eqref{eq:delta_theory}].
In our runs, $\kappa=1/2$ is measured for the post-plateau regime (small~$L/d_e$, large $\tau_{\rm lin}$), and $\kappa=3/2$ for the pre-plateau regime (large~$L/d_e$, small $\tau_{\rm lin}$).
In the asymptotic regime, we expect the scaling $\kappa=2$ based on the expansion of the analytical solution at asymptotically small $tv_{\rm th}/L$ [Eq.~\eqref{eq:Delta_2nd}].

The growth rate of the most unstable mode and its wavenumber in the linear Weibel stage is expected to have power-law dependencies on anisotropy: $\gamma_B \simeq \Delta^{\alpha} \omega_{\rm p} v_{\rm th}/c$ [Eq.~\eqref{eq:gammab}] and $k_{\rm w}d_e \simeq \Delta^{\nu}$ [Eq.~\eqref{eq:kw_delta}].
Fig.~\ref{fig:weibel_gamma_kw} shows the measured magnetic growth rate at $\tau_{\rm lin}$, $\gamma_{B,{\rm max}}$, (top panel) and the normalized wavenumber, $k_{\rm w}d_e$, corresponding to the peak of the isotropic magnetic power spectrum $M(k)$ at $\tau_{\rm lin}$ (bottom panel), as functions of measured $\Delta_{\rm max}$ for runs with varying $L/d_e$.
The $\gamma_{B,{\rm max}}/\omega_{{\rm p}} \propto \Delta_{\rm max}^2$ (i.e., $\alpha=2$) and $k_{\rm w}d_e \propto \Delta_{\rm max}$ (i.e., $\nu=1$) scalings are found across most of the values of~$L/d_e$, except for the two runs with the largest $L/d_e$ (corresponding to the two data points on the left with the smallest~$\Delta_{\rm max}$). 
These measured scalings are different from the expected scalings ($\alpha=3/2$ and $\nu=1/2$) for the asymptotic regime and from the canonical Weibel theory~\cite{davidson1972nonlinear}.

In the same figure, we plot with the brown dash-dotted lines the analytical growth rate of the most unstable Weibel mode (top panel) and its corresponding wavenumber (bottom panel) as functions of pressure anisotropy,  given by the asymptotic solution of the linear Weibel dispersion relation (see the Supplementary materials for detailed derivation).
This solution is obtained in the regime where an asymptotically large scale separation exists.
With a large enough $L/d_e$ (the two runs with $L/d_e=768,1024$), the measured growth rate and wavenumber agree well with the asymptotic solution, confirming that the primary instability producing the magnetic fields in our system is indeed the Weibel instability.
As $L/d_e$ decreases, however, the measured quantities deviate from the asymptotic solution and exhibit different scalings.
We believe that this discrepancy is due to the effects of the continuous forcing under insufficient scale separation ($L/d_e$).
With a limited $L/d_e$, the distribution function is already driven to a complex form when the Weibel instability becomes active (very different from a tri-Maxwellian in the asymptotic regime in an orthonormal coordinate system).
In addition, during the linear Weibel stage, the assumption of a static background is no longer a good approximation if the fluid time scale $L/v_{\rm th}$ is not asymptotically large compared to the inverse growth rate~$1/\gamma_B$; the effect of the shear flow in tilting the Weibel filaments is not negligible.
The combination of these effects leads to different values of Weibel growth rate and wavenumber and their different scaling dependencies on $\Delta$ for limited~$L/d_e$.

The increasing magnetic growth rate leads to super-exponential growth of magnetic energy, and thus of $\beta^{-1}$ [Eq.~\eqref{eq:exponential_B_theory}].
When the argument of the exponential function becomes of order unity, the linear stage ends. This moment corresponds to the measured $\tau_{\rm lin}$. 
This is consistent with the fact that $\beta^{-1}$ in runs with varying $L/d_e$ or $S_0$ reaches the same value at $\tau_{\rm lin}$ (shown by the horizontal dashed lines in bottom panels of each column in Fig.~\ref{fig:scanLdeS0}).

\begin{figure}
    \centering
    \includegraphics[width=0.45\textwidth]{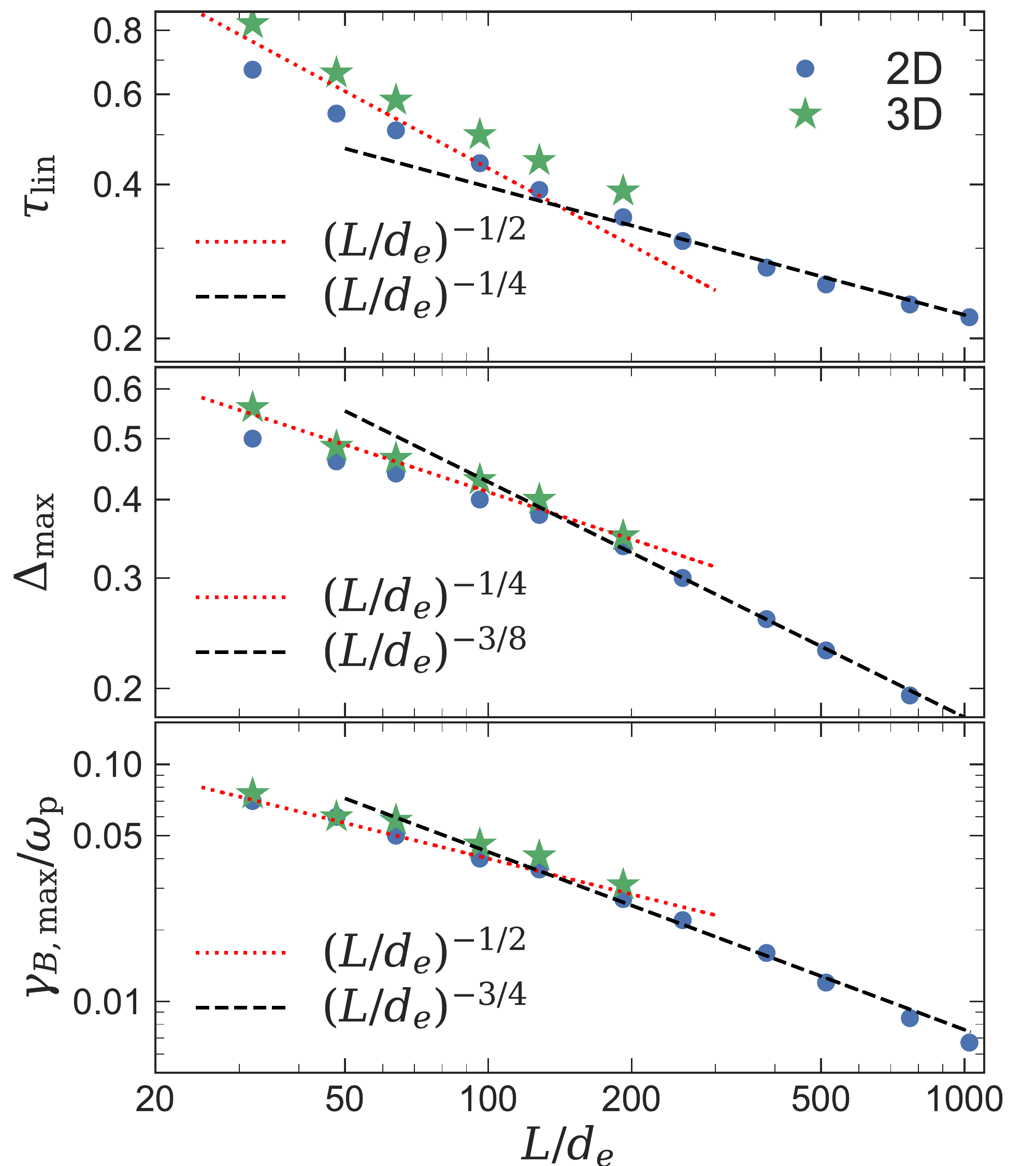}
    \caption{Plots of $\tau_{\rm lin}$ (top), $\Delta_{\rm max}$ (middle), and $\gamma_{B,{\rm max}}/\omega_{\rm p}$ (bottom) versus $L/d_e$ for 2D and 3D runs with varying $L/d_e$ and fixed $S_0=0.2$. Red (black) dotted lines show the predicted scalings in post-plateau (pre-plateau) regime.}
    \label{fig:scaling_Lde_taulin}
\end{figure}

The values of $\tau_{\rm lin}$ and quantities measured at $\tau_{\rm lin}$ are expected to exhibit power-law dependencies on $L/d_e$ and $S_0$, according to Eqs.~\eqref{eq:tau_lin_scaling}--\eqref{eq:gammaB_lin_scaling}.
The exponents $\alpha$ and $\kappa$ are obtained from our numerical results for small and moderate $L/d_e$ (Fig.~\ref{fig:detla_taulin}), and are obtained from the analytical solution at $tv_{\rm th}/L \ll 1$ for asymptotically large $L/d_e$ [Eqs.~\eqref{eq:Us_2nd} and \eqref{eq:Delta_2nd}].
Plugging the measured values $\alpha=2$ and $\kappa \in \{1/2,\ 3/2\}$ into Eqs.~\eqref{eq:tau_lin_scaling}--\eqref{eq:gammaB_lin_scaling}, we derive the following scalings: for the $L/d_e$ dependence, we expect that in the post-plateau regime ($\kappa=1/2$), $\tau_{\rm lin} \sim (L/d_e)^{-1/2}$, $\Delta_{\rm max} \sim (L/d_e)^{-1/4}$, and $\gamma_{B,{\rm max}} \sim (L/d_e)^{-1/2}$; in the pre-plateau regime ($\kappa=3/2$), $\tau_{\rm lin} \sim (L/d_e)^{-1/4}$, $\Delta_{\rm max} \sim (L/d_e)^{-3/8}$, and $\gamma_{B,{\rm max}} \sim (L/d_e)^{-3/4}$.
These latter (pre-plateau) scalings are close to those in the asymptotic regime, for which we expect  $\tau_{\rm lin} \sim (L/d_e)^{-1/4}$, $\Delta_{\rm max} \sim (L/d_e)^{-1/2}$, and $\gamma_{B,{\rm max}} \sim (L/d_e)^{-3/4}$ [Eqs.~\eqref{eq:tau_lin_scaling_largeLde}--\eqref{eq:gammaB_lin_scaling_largeLde}].
The above predicted scalings for the post- and pre-plateau regimes are confirmed by the numerical results shown in Fig.~\ref{fig:scaling_Lde_taulin}, where the transition of scalings occurs at around $L/d_e \approx 200$, consistent with what we observe in Fig.~\ref{fig:scanLdeS0}. 

\begin{figure}[h!]
    \centering
    \includegraphics[width=0.45\textwidth]{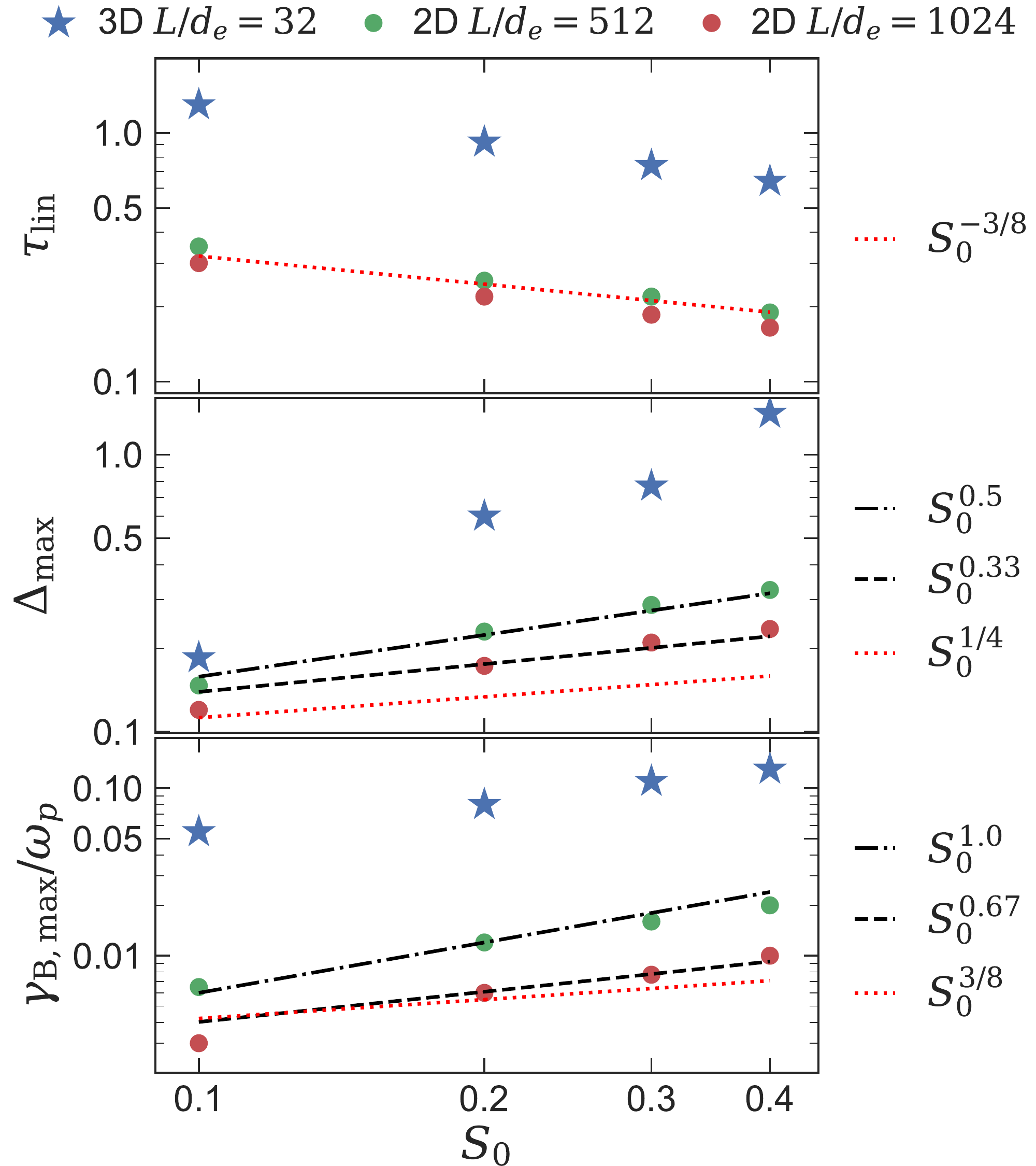}
    \caption{Plots of $\tau_{\rm lin}$ (top), $\Delta_{\rm max}$ (middle), and $\gamma_{B,{\rm max}}/\omega_{\rm p}$ (bottom) versus $S_0$ for 2D and 3D runs with varying~$S_0$. Red dotted lines show the theoretical predictions and black dashed lines show fits to the data points. With increasing $L/d_e$, the measured scalings approach the predictions.}
    \label{fig:scaling_S0_taulin}
\end{figure}

The dependence of $\tau_{\rm lin}$, $\Delta_{\rm max}$, and $\gamma_{\rm B, max}$ on $S_0$ ($\hat{a}_0 \propto S_0$) is more difficult to test in our numerical results. 
For runs with varying $S_0$, the background evolution of $M$ and $\Delta$ for the unmagnetised plasma differs and the transition between the pre- and post-plateau regimes occurs at different critical values of~$L/d_e$.
For fixed small or moderate $L/d_e$, $\Delta$ scales differently with time (at around~$\tau_{\rm lin}$) for systems with different~$S_0$, rendering the application of our scaling theory nontrivial.
We therefore focus on the regime with asymptotically large $L/d_e$, where the quadratic time dependence of $\Delta$ [Eq.~\eqref{eq:Delta_2nd}] applies to systems with any values of~$S_0$.
In this asymptotic regime, quantities are expected to scale with $S_0$ as $\tau_{\rm lin} \sim S_0^{-3/8}$, $\Delta_{\rm max} \sim S_0^{1/4}$, $\gamma_{B,{\rm max}}/\omega_{\rm p} \sim S_0^{3/8}$ [Eqs.~\eqref{eq:tau_lin_scaling_largeLde}-\eqref{eq:gammaB_lin_scaling_largeLde}], shown by the red dotted lines in Fig.~\ref{fig:scaling_S0_taulin}.
Three groups of runs with different values of $L/d_e$ fixed in each case and with a parameter scan on $S_0$ are presented.
We are not able to perform simulations deep in the asymptotic regime due to computational constraints, especially in~3D.
However, it seems clear that with increasing $L/d_e$ the measured scalings approach our asymptotic predictions.

\subsubsection{Scaling laws at the saturation of Weibel instability.}

\begin{figure}
    \centering
    \includegraphics[width=0.45\textwidth]{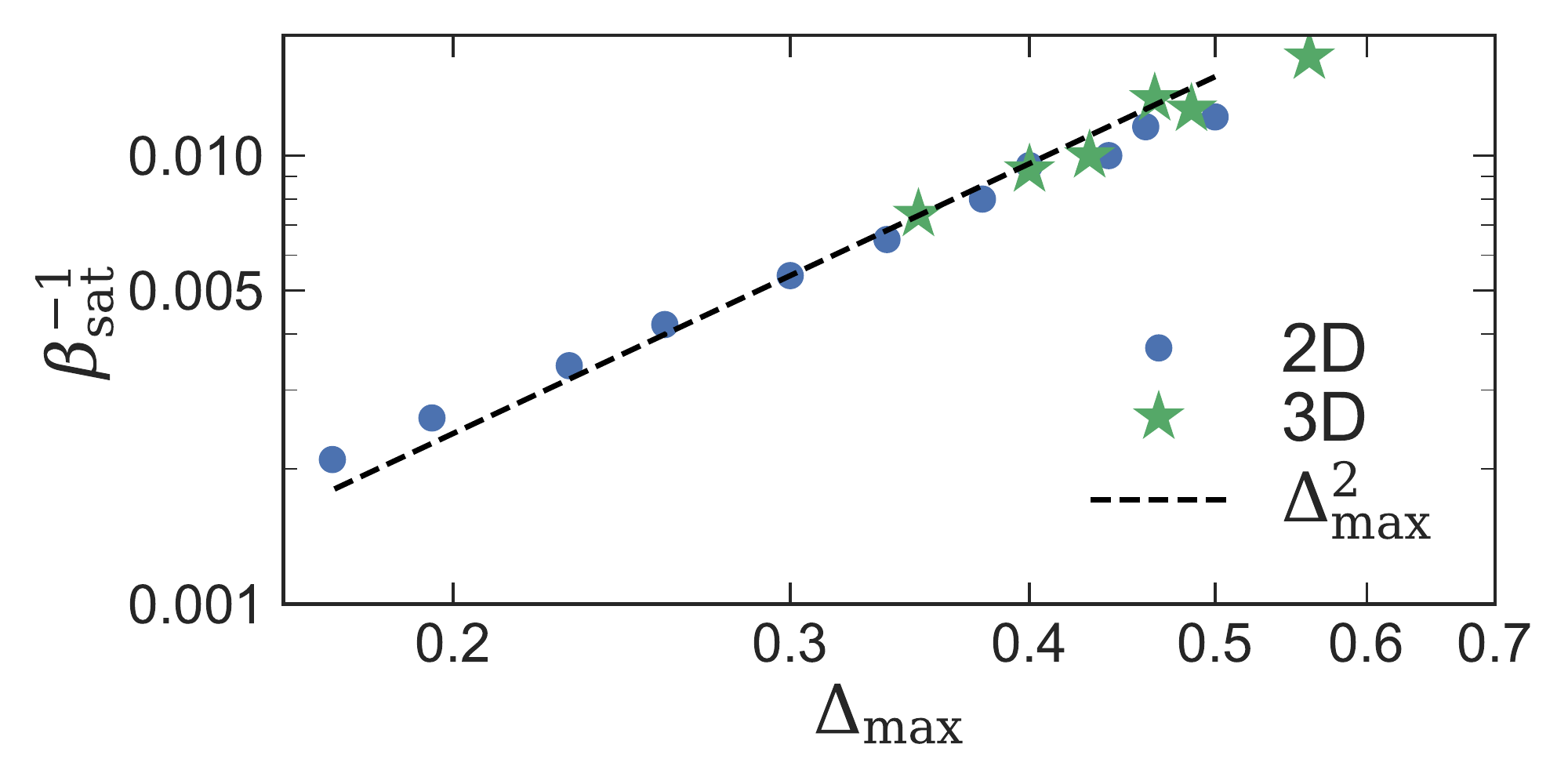}
    \includegraphics[width=0.45\textwidth]{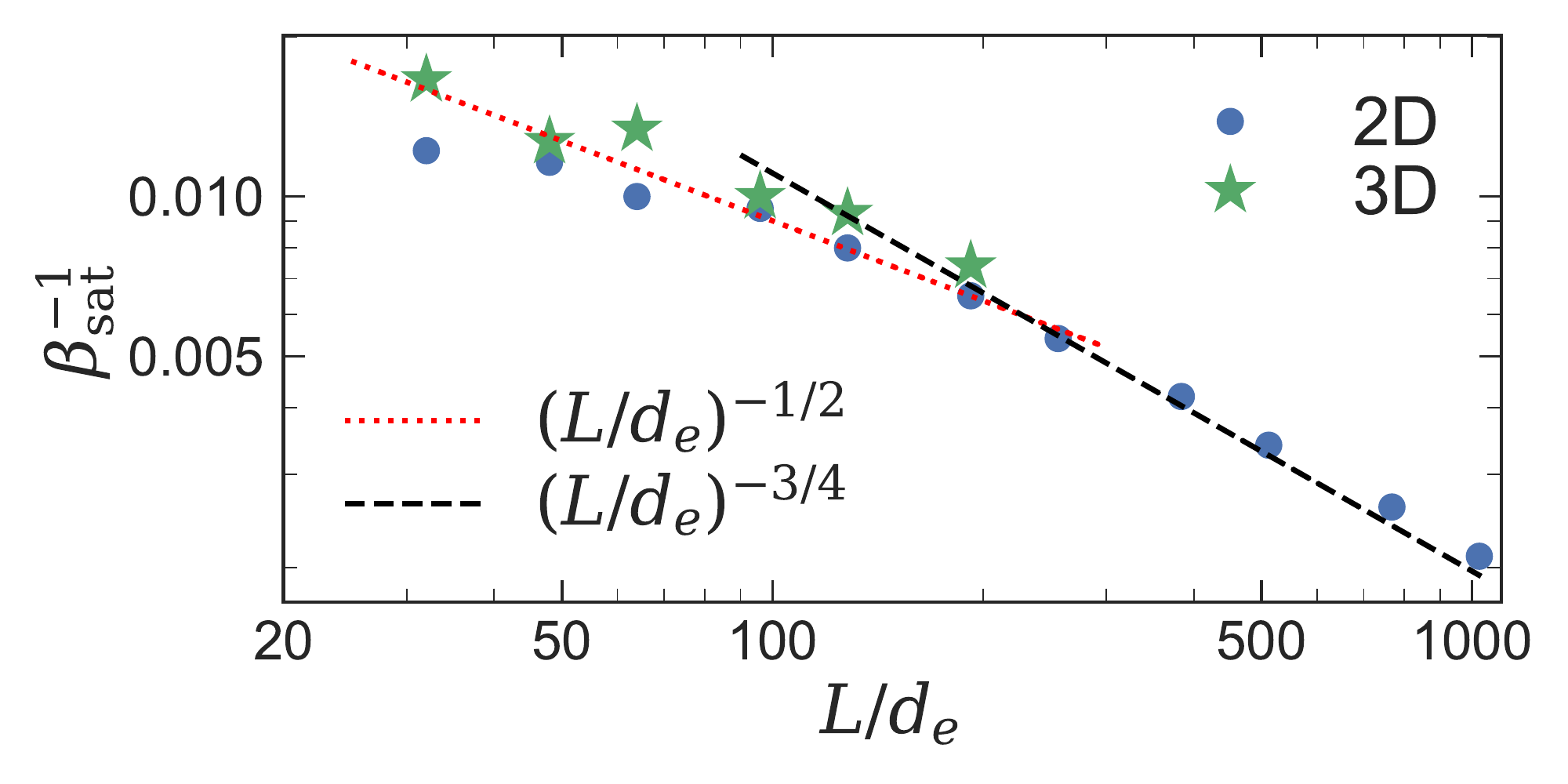}
    \caption{Saturated inverse beta $\beta_{\rm sat}^{-1}$ versus $\Delta_{\rm max}$ (top) and versus $L/d_e$ (bottom) for 2D and 3D runs with varying $L/d_e$ and fixed $S_0=0.2$. }
    \label{fig:scaling_Lde_tausat}
\end{figure}

\begin{figure}
    \centering
    \includegraphics[width=0.45\textwidth]{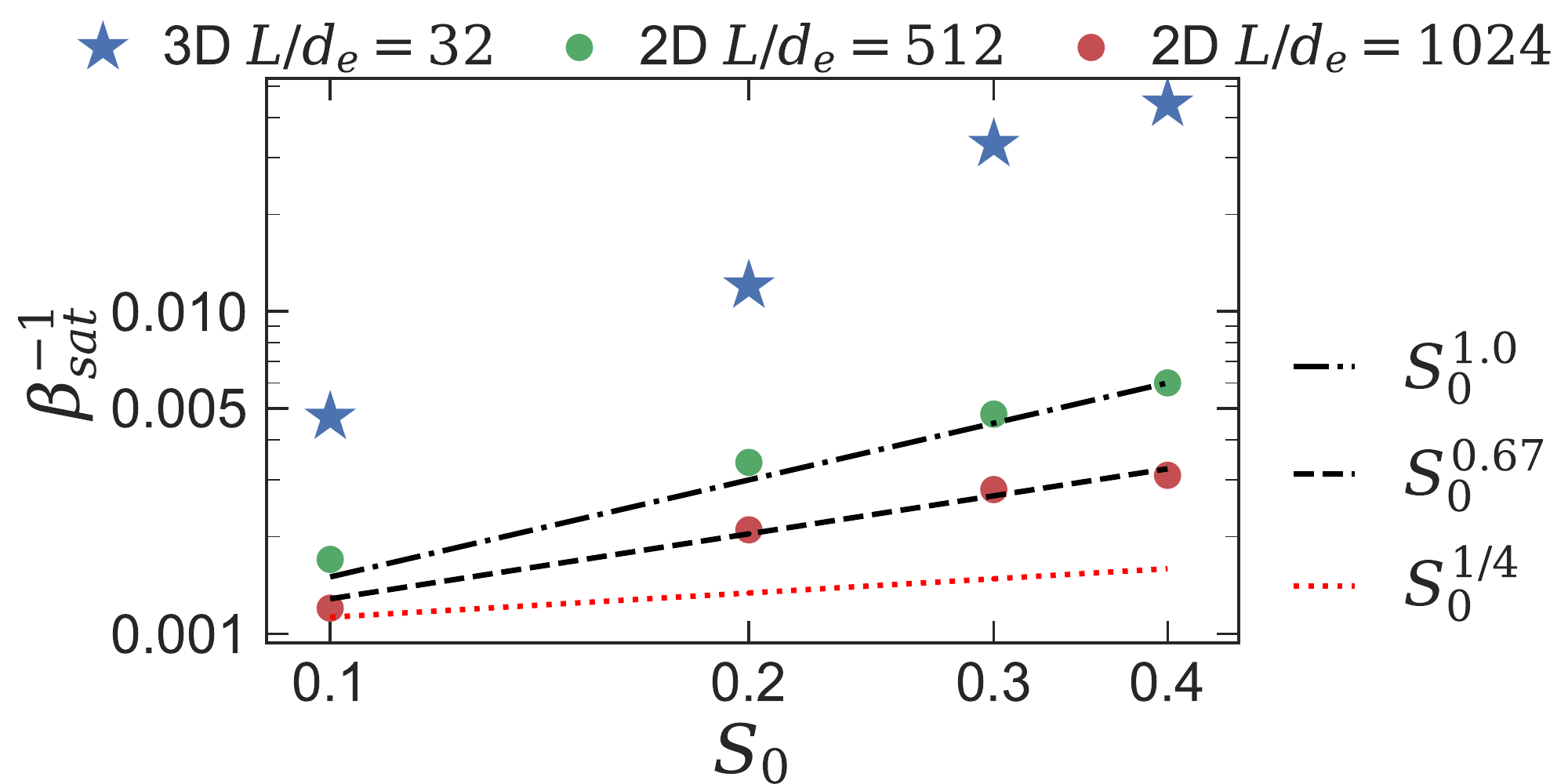}
    \caption{Saturated inverse beta $\beta_{\rm sat}^{-1}$ versus $S_0$ for 2D and 3D runs with varying~$S_0$. }
    \label{fig:scaling_S0_tausat}
\end{figure}

The saturation of Weibel instability that we observe in Sec.~\ref{sec:sim_single_saturation} occurs when the produced magnetic fields become strong enough to instigate particles' gyromotion on the length scale of magnetic filaments, i.e.,~$k_{\rm w} \rho_e \sim 1$~\cite{davidson1972nonlinear,kato2005saturation}. 
As discussed in Sec.~\ref{sec:theory_saturation}, at saturation, $\rho_e$ is related to the saturated magnetic field as~$\rho_e \simeq \beta_{\rm sat}^{1/2}d_e$, and $k_{\rm w}$ is approximated with the inverse length scale of the magnetic field at $\tau_{\rm lin}$, determined by~$\Delta_{\rm max}$: $k_{\rm w}(\tau_{\rm lin})\simeq \Delta_{\rm max}^\nu/d_e$ [Eq.~\eqref{eq:kw_delta}].
The index $\nu=1$ is measured for the post- and pre-plateau regimes (Fig.~\ref{fig:weibel_gamma_kw}, bottom panel), while $\nu=1/2$ is expected for the asymptotic regime.
The scaling $\beta_{\rm sat}^{-1} \sim \Delta_{\rm max}^2$ [Eq.~\eqref{eq:betasat_Delta}] immediately follows (with $\nu=1$), and is confirmed both in the post- and pre-plateau regimes (Fig.~\ref{fig:scaling_Lde_tausat} top panel).
Combined with the dependence of $\Delta_{\rm max}$ on $L/d_e$ and $S_0$ [Eqs.~\eqref{eq:delta_lin_scaling} and \eqref{eq:delta_lin_scaling_largeLde}], we obtain the following predictions [Eqs.~\eqref{eq:betasat_Lde} and \eqref{eq:betasat_Lde_largeLde}]: in the post-plateau regime, $\beta_{\rm sat}^{-1} \sim (L/d_e)^{-1/2}$; in the pre-plateau regime $\beta_{\rm sat}^{-1} \sim (L/d_e)^{-3/4}$; and in the asymptotic regime, $\beta_{\rm sat}^{-1} \sim (L/d_e)^{-1/2}$.
The scalings in the post- and pre-plateau regimes are confirmed by the numerical results (Fig.~\ref{fig:scaling_Lde_tausat}, bottom panel).
For the same reason explained in Sec.~\ref{sec:sim_sclaing_linear}, we are only able to predict the dependence of $\beta_{\rm sat}^{-1}$ on $S_0$ for systems with asymptotically large $L/d_e$: $\beta_{\rm sat}^{-1} \sim S_0^{1/4}$ [Eq.~\eqref{eq:betasat_Lde_largeLde}]. 
Although we are not able to perform simulations deep in this asymptotic regime, a clear trend is shown in Fig.~\ref{fig:scaling_S0_tausat} that the measured scalings approach the $S_0^{1/4}$ prediction with increasing $L/d_e$.\\

The presented numerical results confirm our analytical model (Sec.~\ref{sec:theory}) in the post- and pre-plateau regimes (for small and moderate~$L/d_e$).
The three exponents in the model, $\alpha$, $\kappa$, and $\nu$, are determined by the numerical results. 
The derived scalings [Eqs.~\eqref{eq:tau_lin_scaling}--\eqref{eq:gammaB_lin_scaling} and \eqref{eq:betasat_Delta}--\eqref{eq:betasat_Lde}], whose indices are functions of $\alpha$, $\kappa$, and $\nu$, are confirmed independently by the numerical results.
The validation of our model in the post- and pre-plateau regimes gives us confidence in its predictions in the asymptotic regime, which are derived within the same framework as the other regimes. 
As indicated in Fig.~\ref{fig:weibel_gamma_kw}, the two largest 2D runs ($L/d_e$=768, 1024) marginally enter the asymptotic regime.
However, in order to enter the deep asymptotic regime and obtain the relevant scalings, the short-time ($tv_{\rm th}/L \lesssim 0.1$) approximation of the unmagnetized solution [Eq.~\eqref{eq:fs_2nd_order}] needs to be valid during the growth of Weibel seed fields (at $tv_{\rm th}/L \simeq \tau_{\rm lin}$), i.e., $\tau_{\rm lin} \lesssim 0.1$.
The weak scaling dependence $\tau_{\rm lin} \sim (L/d_e)^{-1/4}$ [Eq.~\eqref{eq:tau_lin_scaling_largeLde}] then suggests that a significantly larger scale separation, $L/d_e \gtrsim 10^4$, is required to assess the deep asymptotic regime.


\section{Discussion}
\label{sec:discussion}

This paper provides a clear demonstration and quantitative description of the spontaneous magnetization of collisionless plasma under the action of a shear flow.
The primary kinetic instability that produces the seed magnetic fields is identified as the Weibel instability. 
We predict that in the regime with an asymptotically large time- and length-scale separation, quantified by $L/d_e$ (ratio of system scale to electron skin depth), the saturated seed magnetic fields are expected to have a characteristic length scale $\lambda_{\rm w} \sim L^{1/4}d_e^{3/4}\hat{a}_0^{-1/8}$~[Eq.~\eqref{eq:lambdaw_Lde_asym}], and lead to a saturated inverse beta $\beta_{e,{\rm sat}}^{-1} \sim (L/d_e)^{-1/2}\hat{a}_0^{1/4}$ [Eq.~\eqref{eq:betasat_Lde_largeLde}], where $\hat{a}_0$ is the normalized acceleration driving the macroscopic shear flow. 
The relatively weak $(L/d_e)^{-1/2}$ dependence of $\beta^{-1}_{e,{\rm sat}}$ implies that in very large astrophysical systems the Weibel instability generates much stronger seed fields than those thought to be produced by the Biermann battery, for which $\beta^{-1}_{\rm sat}\propto L^{-2}$~\cite{max1978enhanced,haines1997saturation}. After saturation, the Weibel filaments undergo long-term evolution that sees their scale increase gradually towards the system size through coalescence. 

The Weibel instability has been historically analyzed within context of counter-streaming flows~\cite{silva2003interpenetrating,fox2013filamentation,huntington2015observation} or collisionless shocks~\cite{medvedev1999generation,Kato2008,medvedev2006cluster,spitkovsky2008}, in which the external drive is strong and/or the Mach number is typically high. 
In this work, however, we consider a weakly driven, large-scale shear flow.
This constitutes an important step in establishing a connection with a broader set of astrophysical applications beyond shock physics, such as low-Mach-number turbulence in galaxy clusters and in the intergalactic media.
The production of magnetic fields has also been studied in the configuration of counter-streaming flows through the kinetic Kelvin-Helmholtz instability~\cite{alves2012large,alves2014electron,grismayer2013dc,nishikawa2014magnetic} and of differential rotation through electron instabilities~\cite{quataert2015linear,heinemann2014linear}. 
By contrast, rather than initialize a configuration that is super-critical to the instabilities of interest, we instead start with a stable equilibrium and drive the system gradually towards becoming marginally unstable to the relevant kinetic instability (in this case, the Weibel).

It is important to note that ion kinetic physics is not taken into account in this study. With the development of ion thermal pressure anisotropy, the ion Weibel instability can in principle also be triggered and produce seed magnetic fields on a time scale ${\sim}\omega^{-1}_{{\rm p}i}$ and a length scale ${\sim}d_i$, where $d_i=c/\omega_{{\rm p}i}$ is the ion skin depth.   
However, when the ion Weibel instability becomes active, the electron Weibel instability should already be saturated and the electrons already magnetized. 
In this high-$\beta$ system, various electron-pressure-anisotropy instabilities are expected play a role~(e.g., electron firehose, whistler; \cite{riquelme2016}), and it is not clear how these electron-scale instabilities might interplay with the ion Weibel instability (or, for that matter, subsequent ion-pressure-anisotropy instabilities like firehose and mirror~\cite{kunz2014firehose,riquelme2015,riquelme2018}). 
We defer the inclusion of ion kinetic physics to future work.

In the meantime, it is worth applying our results to an actual astrophysical system, if only suggestively. For example, in the hot and dilute ICM, the scale of observed macroscopic turbulent motions is $L \gtrsim 10~{\rm kpc}$, while the electron skin depth may be estimated from the observed electron density as $d_e \sim 10^{-12}~{\rm pc}$. 
This gives a typical scale separation of $L/d_e \gtrsim 10^{16}$. For this ratio,  Eq.~\eqref{eq:betasat_Lde_largeLde} leads us to expect the saturated seed magnetic fields produced by the electron Weibel instability to give $\beta_{\rm sat} \sim (L/d_e)^{1/2} \sim 10^{8}$. Under typical cluster conditions, this value of $\beta_{\rm sat}$ corresponds to a ${\sim}0.1~{\rm nG}$ magnetic field. 
Despite the relatively small scale of this field, its amplitude is notable because the Weibel's main competitor, the Biermann battery, produces fields that are much weaker, at ${\sim}10^{-20}~{\rm G}$~\cite{biermann1950biermann,kulsrud1997protogalactic}.
Interestingly, configurations that give rise to Biermann fields---misalignment of plasma density and pressure gradients---have been shown to be unstable to the Weibel instability as well; the ensuing strong small-scale seed fields are radically different from their more conventional Biermann origin~\cite{schoeffler2014magnetic,schoeffler2018fully}.

Despite their initially small (electron) scales, we argue that the saturated Weibel seed fields---whose morphology is that of flux ropes---can inverse-cascade to larger scales through magnetic reconnection~\cite{medvedev2004long,zhou2019magnetic,zhou2020multi,bhat2021inverse,zhou2021statistical,hosking2020reconnection}. 
This inverse cascade should reach the scale at which the reconnection time scale of the seed fields becomes comparable to the nonlinear eddy turn-over time scale of the turbulent flow. 
Above this critical scale, the coalescence of Weibel seed fields may be expected to be replaced by amplification of those fields through the turbulent dynamo. The feasibility of this scenario will be addressed in a separate publication.

This work provides the first step in the building of a new paradigm for understanding magnetogenesis in the Universe. 
It quantitatively describes the emergence and evolution of seed magnetic fields that arise self-consistently from generic motions (shear flows) that can also support a turbulent dynamo. 
Future investigations are required to determine how such seed fields can be amplified by astrophysical turbulence to dynamically important levels on cosmologically short times scales~\cite{zhou2020weibel}. 
This new paradigm will provide a fully self-consistent explanation for the origin and prevalence of cosmic magnetism---one of the most important science drivers of upcoming radio telescopes such as the Square Kilometer Array.

\paragraph{Acknowledgements.}
The authors thank J.~Juno, F.~Rincon, A.~A.~Schekochihin, and D.~A.~St-Onge for insightful discussions.
Support for NFL and MZ was provided by the National Science Foundation (NSF) under CAREER award No.~1654168 and by the National Aeronautics and Space Administration (NASA) under award NNH19ZA001N-FINESST. Support for VZ was provided by NASA Hubble Fellowship grant \#HST-HF2-51426.001-A awarded by the Space Telescope Science Institute, which is operated by the Association of Universities for Research in Astronomy, Inc., for NASA, under contract NAS5-26555. Support for MK was provided by NSF CAREER award No.~1944972. Support for DAU was provided by NASA grants NNX17AK57G and 80NSSC20K0545, and NSF grant AST-1806084.
The completion of this work was aided by the generous hospitality of the Kavli Institute for Theoretical Physics in Santa Barbara, supported in part by the National Science Foundation under Grant No.~NSF PHY-1748958.
The work used the Extreme Science and Engineering Discovery Environment (XSEDE), which is supported by National Science Foundation Grant No.~ACI-1548562. The simulations presented in this work were performed on the supercomputer Stampede2 at the Texas Advanced Computer Center (TACC) through allocation No.~TG-PHY140041~\cite{towns2014xsede}.

\bibliography{ref}

\clearpage
\appendix

\setcounter{equation}{0}
\setcounter{figure}{0}
\setcounter{table}{0}
\newcounter{SIfig}

\renewcommand{\theequation}{S\arabic{equation}}
\renewcommand{\thefigure}{S\arabic{figure}}
\renewcommand{\thetable}{S\arabic{table}}
\renewcommand{\theSIfig}{S\arabic{SIfig}}

\section*{Supplementary Material}

As a supplement to the main text, we calculate here the dispersion relation of the Weibel modes using the unmagnetized solution of the plasma distribution function $f_s$, Eq.~\eqref{eq:f_exact_uncharged_theory}. We first show that $f_s$ is a multivariate distribution function under certain approximations, and can thus be written as a tri-Maxwellian in an orthonormal coordinate system. We then numerically solve the dispersion relation for an oblique Weibel mode in a tri-Maxwellian plasma and find the dependence of the growth rate of the most unstable mode, $\gamma_{\rm w}$, on the thermal pressure anisotropy, $\Delta$. 

\subsection{Coordinate transformation of $f_s$}
\label{sec:f0_transformation}

Let us specify a location $x=0$ at which maximum shear occurs and thereby remove the spatial dependence of $f_s$. The plasma at this position undergoes the strongest phase mixing, and thus has the maximum thermal pressure anisotropy. The  dynamics of the Weibel instability at this position is therefore representative of that in the whole system. In the small-time limit $\epsilon \equiv tv_{\text{th}s}/L \ll 1$ and at $x=0$, Eq.~\eqref{eq:f_exact_uncharged_theory} becomes  
\begin{equation}
\begin{split}
    \widetilde{v}_y &\equiv v_y + \frac{La_0}{2\pi v_x}\left[1-\cos\left(\frac{2\pi}{L}v_xt\right)\right]\\
    &\simeq v_y + \hat{a}_0\pi v_x\left(\frac{tv_{\text{th}s}}{L}\right)^2 + \mathcal{O}(\epsilon^3).
\end{split}
\end{equation}
Combining the time evolution of thermal pressure anisotropy [Eq.~\eqref{eq:Delta_x_2nd}], 
\begin{equation}
    \Delta_s(t,x=0) = \frac{3}{2}\pi\hat{a}_0\left(\frac{tv_{\text{th}s}}{L}\right)^2 + \mathcal{O}(\epsilon^3),
\end{equation}
we can simplify the expression of $\widetilde{v}_y$ as
\begin{equation}
    \widetilde{v}_y \equiv v_y + \frac{2}{3}\Delta_s(t)v_x,
\end{equation}
and that of $f_s$ at $x=0$ as
\begin{equation}
    f_s(\bb{v}) = F_{{\rm M},s}\left[ \biggl( 1 + \frac{4}{9} \Delta^2_s\biggr) v_x^2 + \frac{4}{3}\Delta_s v_x v_y + v_y^2 + v_z^2\right].
\label{eq:fx_x0}
\end{equation}
In this case, $f_s$ possesses the form of a multivariate normal distribution and can thus be transformed to an orthonormal coordinate basis $\{ v_{x'}, v_{y'}, v_z\}$ and written as the tri-Maxwellian distribution
\begin{equation}     \label{eq:multivariate_normal}
    \widetilde{f_s} \propto \exp\left[ -\left( \frac{v_{x'}^2}{2 T_{x',s}} + \frac{v_{y'}^2}{2T_{y',s}} + \frac{v_z^2}{2T_{z,s}} \right) \right].
\end{equation}
Here $T_{x',s}$, $T_{y',s}$, and $T_{z,s}$, with  $T_{y',s}>T_{z,s}>T_{x',s}$, are the eigenvalues of the covariance matrix of $f_s$, and $v_{x'}$, $v_{y'}$, and $v_z$ are the corresponding eigenvectors.
Note that the orientation of the orthonormal coordinate evolves with time.
The thermal pressure anisotropy (defined in Sec.~\ref{sec:theory}) thus becomes $\Delta_s \equiv \sqrt{ \langle (P_{{\rm max},s}/P_{\perp,s})^2 \rangle}-1 = \sqrt{ \langle (T_{y',s}/T_{\perp,s})^2 \rangle}-1$, where $T_{\perp, s} = (T_{x',s}+T_{z,s})/2$. 

\subsection{General dispersion relation for Weibel instability}
We proceed to derive the linear dispersion relation of the oblique Weibel modes for a tri-Maxwellian distribution function. 
The goal of this calculation is to obtain the dependence on pressure anisotropy of the growth rate of the most unstable Weibel mode.
For simplicity, we consider a system that is 3D in velocity space ($v_{x'}$,$v_{y'}$,$v_z$) and 2D in configuration space ($x'$,$y'$). 
Our numerical results in Sec.~\ref{sec:numerical_results} show that, at least for the unmagnetized stage and the linear Weibel stage, systems with 3D and 2D configuration space exhibit almost identical results, thereby justifying this approximation. 

We begin by considering the tri-Maxwellian initial distribution 
\begin{equation}\label{eqn:triMax}
    \widetilde{f}_{0,s}(v_{x'}, v_{y'}, v_z) = \widetilde{f}_{0x',s}(v_{x'})\widetilde{f}_{0y',s}(v_{y'})\widetilde{f}_{0z,s}(v_z),
\end{equation}
where
\begin{equation}
    \widetilde{f}_{0a,s}(v_a) = \frac{1}{\sqrt{\pi}v_{{\rm th} a,s}}\exp\left\{-\frac{v_a^2}{2v_{{\rm th}a,s}^2}\right\},
\end{equation}
$v_{{\rm th}a,s}\equiv\sqrt{T_{a,s}/m_s}$, and $a \in \{x',y',z\}$. To this distribution we add a linear  perturbation, whose 2D spatial dependence is characterized by a wavenumber that contains both transverse and longitudinal components:
\begin{equation}
    \bb{k} =  k_{x'}\hat{\bb{x}}' + k_{y'}\hat{\bb{y}}' .
\end{equation}
The general expression for the components of the dielectric tensor, which specifies the oscillatory response of the plasma, is
\begin{equation}
\begin{split}
    \epsilon_{ab}(\omega,\mathbf{k}) &= \left(1-\sum_s\frac{\omega_{{\rm p}s}^2}{\omega^2}\right)\delta_{ab} \\
    \mbox{} &+ \sum_s \frac{\omega_{{\rm p}s}^2}{\omega^2}\int\rmd^3\bb{v}\ \frac{v_a v_b}{\omega-\bb{k} \bcdot \bb{v}} \, \bb{k}\bcdot\frac{\partial \widetilde{f}_{0,s}}{\partial\bb{v}},
\end{split}
\end{equation}
where $\omega$ is the (complex) frequency of the response. The components of the associated dispersion matrix are given by
\begin{equation}
    D_{ab}(\omega,\bb{k}) = \epsilon_{ab} + \frac{k_ak_b}{\omega^2}c^2 - \frac{k^2c^2}{\omega^2}\delta_{ab},
\end{equation}
where $k=|\bb{k}|$. Plugging in the tri-Maxwellian distribution function $\widetilde{f}_{0,s}$ [Eq.~\eqref{eqn:triMax}] and defining the variables $\xi\equiv (\omega-k_{y'}v_{y'})/|k_{x'}|v_{{\rm th} x'} $, $u \equiv v_{x'}/v_{{\rm th} x'}$, and  $\mathcal{Z}(\xi) \equiv \pi^{-1/2}\int\rmd u\,\exp(-u^2)(u-\xi)^{-1}$, we obtain 
\begin{equation}
\begin{split}
    D_{y'y'} &= 1-\frac{k_{x'}^2c^2}{\omega^2} + \sum_s\frac{\omega_{{\rm p}s}^2}{\omega^2}\Bigg\{-1 + \frac{T_{y'}}{T_{x'}} \\ 
    \mbox{} &+\frac{k_{y'}}{k_{x'}}\frac{v_{{\rm th} {y'}}}{v_{{\rm th} {x'}}}\int\rmd v_{y'}\, \frac{v_{y'}^3}{v_{{\rm th}y'}^3} \widetilde{f}_{y'} \mathcal{Z}(\xi) \\
    \mbox{} &+ 2\frac{v^2_{{\rm th}{y'}}}{v^2_{{\rm th}{x'}}}\int\rmd v_{y'}\, \frac{v_{y'}^2}{v^2_{{\rm th}{y'}}}\widetilde{f}_{y'}\xi \mathcal{Z}(\xi)\Bigg\},
\end{split}
\end{equation}
\begin{equation}
\begin{split}
    D_{y'x'} &= \ D_{x'y'} = \frac{k_{y'}k_{x'}c^2}{\omega^2} + \sum_s\frac{\omega_{{\rm p} s}^2}{\omega^2}\Bigg\{\frac{k_{y'}}{k_{x'}}\\ 
    \mbox{} &+ \frac{k_{y'}}{k_{x'}}\int\rmd v_{y'}\, \frac{v_{y'}^2}{v_{{\rm th}y'}^2} \widetilde{f}_{y'} \xi \mathcal{Z}(\xi) \\
    \mbox{} &+ 2\frac{v_{{\rm th} y'}}{v_{{\rm th} x'}}\int\rmd v_{y'}\, \frac{v_{y'}}{v_{{\rm th} x'}}\widetilde{f}_{y'}\xi \bigl[1+\xi \mathcal{Z}(\xi)\bigr]\Bigg\},
\end{split}
\end{equation}
and
\begin{equation}
\begin{split}
    D_{x'x'} &= 1-\frac{k_{y'}^2c^2}{\omega^2} + \sum_s\frac{\omega_{{\rm p}s}^2}{\omega^2}\\ \mbox{} &\times \Bigg\{\frac{k_{y'}}{k_{x'}} \frac{v_{{\rm th} x'}}{v_{{\rm th}y'}}\int\rmd v_{y'}\, \frac{v_{y'}}{v_{{\rm th}y'}} \widetilde{f}_{y'} \xi\bigl[1+\xi \mathcal{Z}(\xi)\bigr] \\
    \mbox{} &+ 2\int\rmd v_{y'}\, \widetilde{f}_{y'}\xi^2 \bigl[1+\xi \mathcal{Z}(\xi)\bigr]\Bigg\}.
\end{split}
\end{equation}
The nontrivial solution of the mode's dispersion relation is given by
\begin{equation}
\label{eq:dispersion_condition}
    \det\msb{D} = 0 \implies D_{y'y'}D_{x'x'}-D_{y'x'}D_{x'y'} = 0 .
\end{equation}

\begin{figure}
    \centering
    \includegraphics[width=0.45\textwidth]{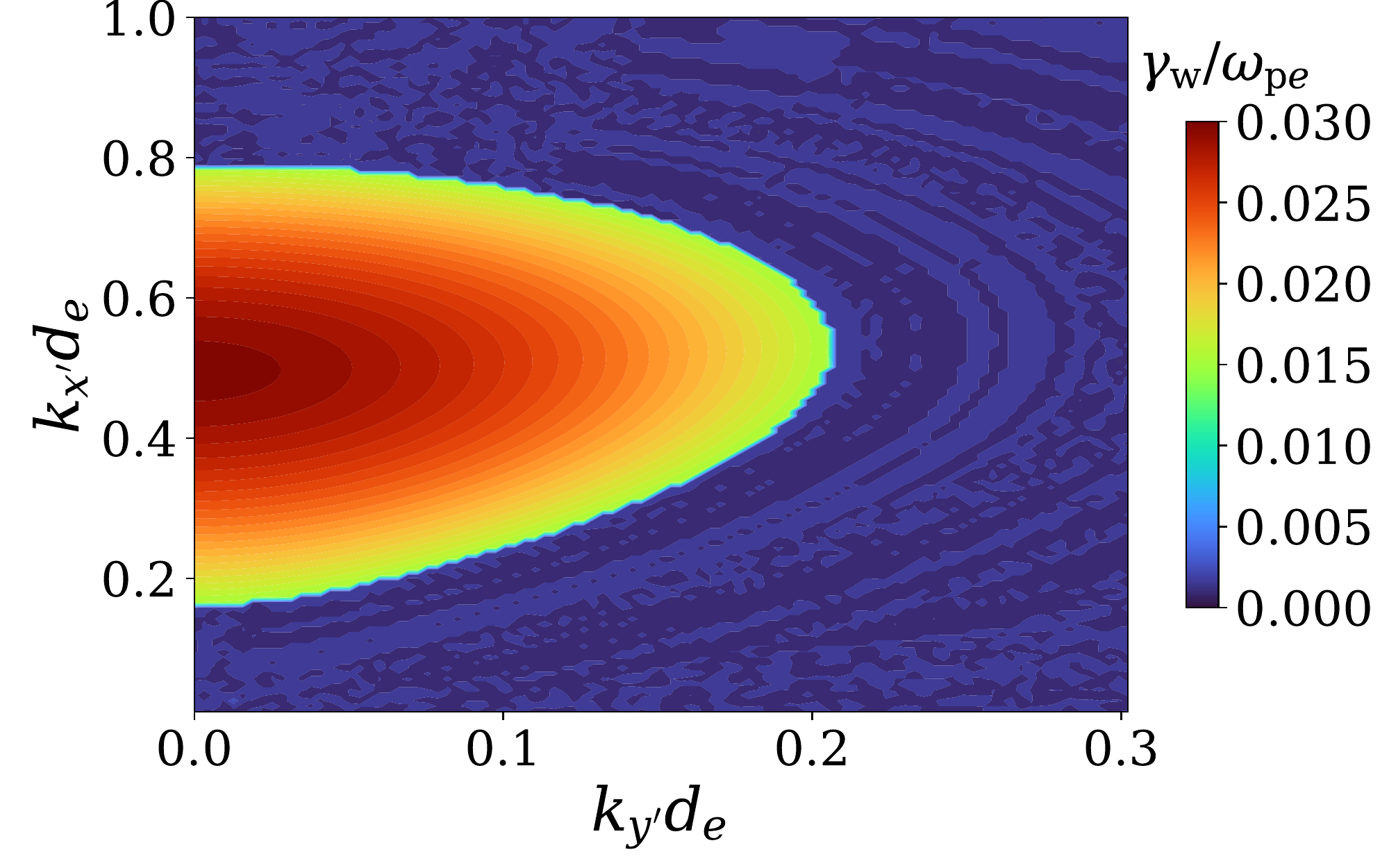}
    \caption{Two-dimensional spectrum of the normalized growth rate of the Weibel modes, $\gamma_{\rm w}/\omega_{{\rm p}e}$, in terms of $k_{x'}d_e$ and $k_{y'} d_e$ for $\Delta_e=0.4$. The most unstable mode is the purely transverse mode ($k_{y'} d_e=0$).}
    \label{fig:gamma_2dmap}
\end{figure}

\begin{figure}
    \centering
    \includegraphics[width=0.4\textwidth]{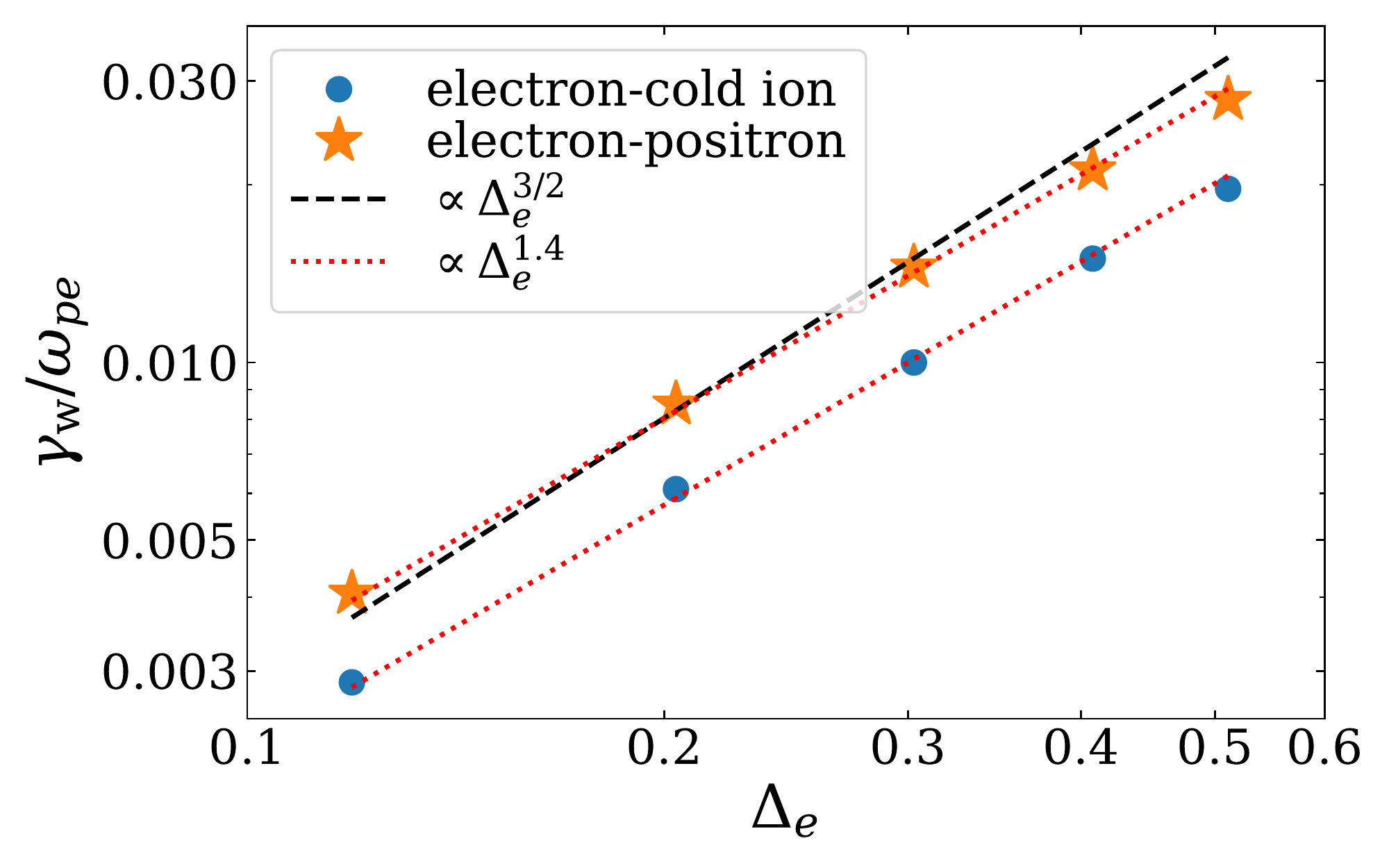}
    \includegraphics[width=0.4\textwidth]{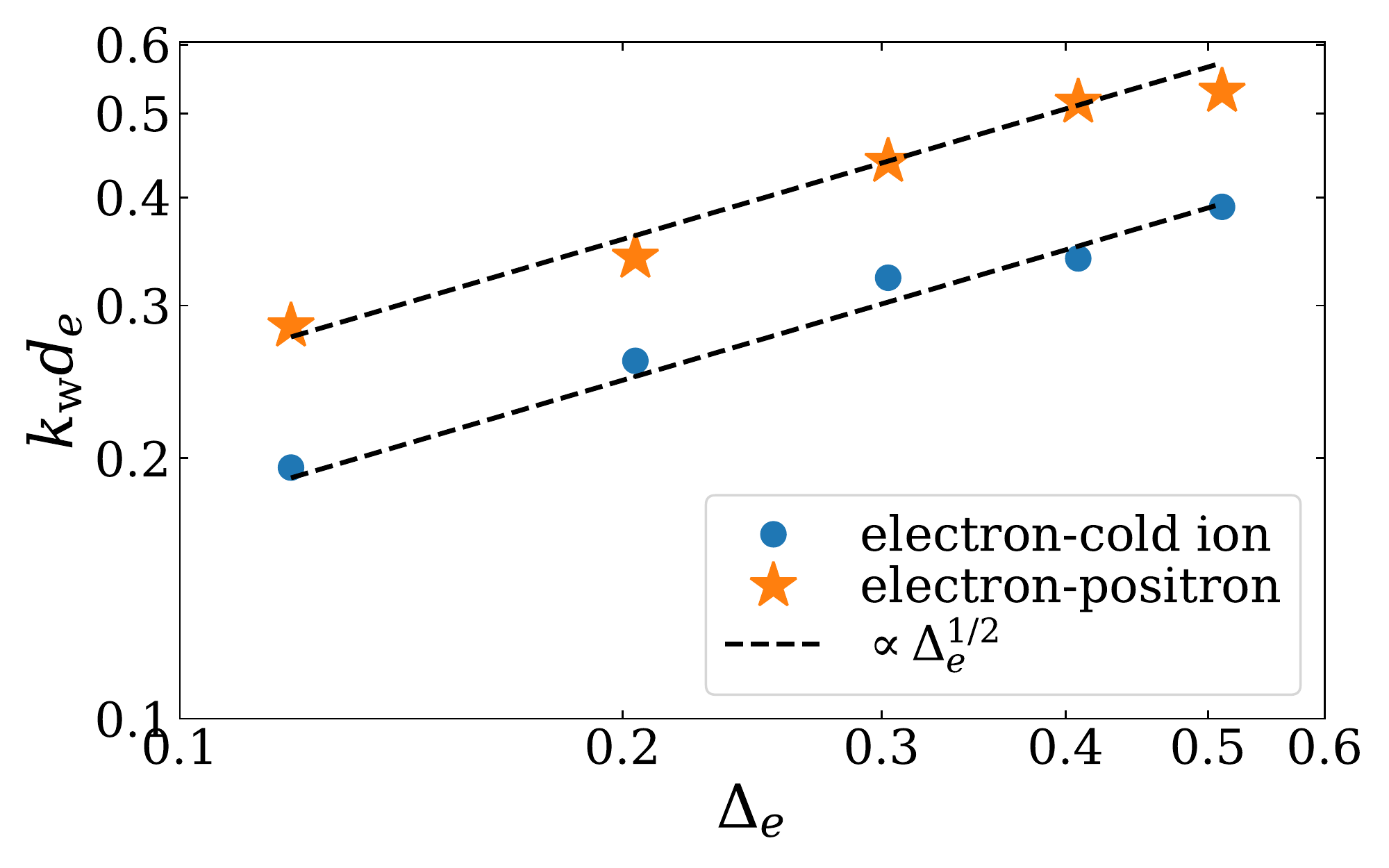}
    \caption{Numerical solution of the Weibel dispersion relation. Top: Maximum normalized Weibel growth rate, $\gamma_{\rm w}/\omega_{{\rm p}e}$, versus the thermal pressure anisotropy. The scalings $\gamma_{\rm w}/\omega_{{\rm p}e} \sim \Delta_e^{3/2}$ and $\gamma_{\rm w}/\omega_{{\rm p}e} \sim \Delta_e^{1.4}$ are shown for reference. Bottom: Normalized wavenumber of the most unstable Weibel mode, $k_{\rm w}d_e$, versus the thermal pressure anisotropy. A $\gamma_{\rm w}/\omega_{{\rm p}e} \sim \Delta_e^{1/2}$ scaling is shown for reference.}
    \label{fig:analytical_gamma_delta}
\end{figure}

We numerically solve Eq.~\eqref{eq:dispersion_condition} for two systems: (i) an electron-positron plasma in which both species respond to the electromagnetic fluctuations and $\Delta_e=\Delta_p$; and (ii) an electron-ion (proton) plasma where only electrons contribute to the Weibel modes and ions are considered as a cold and immobile neutralizing background.  
For a given thermal pressure anisotropy $\Delta_e$, we scan across all~$\bb{k}$ to obtain the 2D spectrum of the Weibel growth rate in terms of $k_{x'} d_e$ and $k_{y'} d_e$.
Fig.~\ref{fig:gamma_2dmap} shows an example for a given $\Delta_e=0.4$.
We find the mode with the largest growth rate $\gamma_{\rm w}$ at the corresponding wavenumber $k_{\rm w}$. 
The dependence of $\gamma_{\rm w}$ and $k_{\rm w}$ on $\Delta_e$ is shown in Fig.~\ref{fig:analytical_gamma_delta}.
The canonical scaling laws $\gamma_{\rm w}/\omega_{{\rm p}e} \sim \Delta_e^{3/2}$ and $k_{\rm w}d_e \sim \Delta_e^{1/2}$~\cite{davidson1972nonlinear} agree well for both a electron-positron plasma and an electron-cold ion plasma.

In addition, we found that the most unstable mode is always the purely transverse mode (i.e., $k_{y'}=0$). This suggests that Weibel instability is the primary instability in the configuration of a driven shear flow at $tv_{\rm th}/L \ll 1$. Other instabilities, such as the electrostatic two-stream instability, do not play a significant role in the system we consider.
This conclusion might be different for other configurations. 
For example, for a system of counter-streaming flows, the dominant instability can be the two-stream instability (especially in the non-relativistic regime), depending on the ratio of flow to thermal velocity~\cite{bret2009weibel}. 

Note that the Weibel growth rate and wavenumber obtained from the dispersion relation~\eqref{eq:dispersion_condition} based on the distribution function in Eq.~\eqref{eq:multivariate_normal}, valid in the small $tv_{\rm th}/L$ limit, is considered as the asymptotic solution. We expect this solution to apply when the system possesses an asymptotically large scale separation~$L/d_e$.

\end{document}